\documentclass[journal]{IEEEtran}
\usepackage{ifpdf}

\usepackage{cite}
\usepackage[cmex10]{amsmath}
\usepackage{algorithmic}

\usepackage{array}

\ifCLASSOPTIONcompsoc
  \usepackage[caption=false,font=normalsize,labelfont=sf,textfont=sf]{subfig}
\else
  \usepackage[caption=false,font=footnotesize]{subfig}
\fi
\usepackage{fixltx2e}
\usepackage{amsfonts}
\usepackage{color}
\usepackage[normalem]{ulem}
\usepackage{bm}

\usepackage{tikz}
\usepackage{pgfplots}
\pgfplotsset{compat=newest}
\usetikzlibrary{plotmarks}
\usepackage{grffile}
\usepackage{multirow}

\newtheorem{theorem}{Theorem}

\newtheorem{remark}{Remark}

\graphicspath{{fig/}}

\makeatletter

\newcommand{\Rmnum}[1]{\expandafter\@slowromancap\romannumeral #1@}

\newcommand{\tabincell}[2]{\begin{tabular}{@{}#1@{}}#2\end{tabular}}

\makeatother

\def\hrulefill{\leavevmode\leaders\hrule height 1.1pt\hfill\kern0pt}
\makeatletter
\def\rulefill{\leavevmode\leaders\hrule depth -3pt height 6pt\hfill\kern0pt}
\makeatother

\pgfplotsset{plot coordinates/math parser=false}
\newlength\figureheight
\newlength\figurewidth

\begin{document}
%
\title{Achieving Energy-Efficient Uplink URLLC with MIMO-Aided Grant-Free Access}

%
%
%

\author{{Linlin~Zhao, Shaoshi~Yang, \emph{Senior Member, IEEE}, Xuefen~Chi, Wanzhong~Chen,  Shaodan~Ma, \emph{Member, IEEE} }
\thanks{Manuscript received October 23, 2020; revised January 13, 2021 and
	May 10, 2021; accepted August 2, 2021. 
	This work is financially supported by the National Natural Science Foundation of China (No. 61801191), the Macau Youth Scholars Program (No. AM201922), the Beijing Municipal Natural Science Foundation (No. L202012), the Open Research Project of the State Key Laboratory of Media Convergence and Communication, Communication University of China (No.
SKLMCC2020KF008), the Fundamental Research Funds for the Central Universities (No. 2020RC05),  the Science and Technology Development Fund, Macau SAR (File
no. 0032/2019/AGJ and File no. SKL-IOTSC-2021-2023), and the Research Committee of University of Macau under Grant MYRG2018-00156-FST.}
\thanks{L. Zhao is with the Department of Communications Engineering, Jilin University, Changchun 130012, China and also with the State Key Laboratory of Internet of Things for Smart City, University of Macau, Macau 999078, China (e-mail: zhaoll13@mails.jlu.edu.cn).}
\thanks{S. Yang (\textit{Corresponding author}) is with the School of Information and Communication Engineering, Beijing University of Posts and Telecommunications, and also with the Key Laboratory of Universal Wireless Communications, Ministry of Education, Beijing 100876, China (e-mail: shaoshi.yang@bupt.edu.cn).}
\thanks{X. Chi and W. Chen are with the Department of Communications Engineering, Jilin University, Changchun 130012, China (e-mail: chixf@jlu.edu.cn, chenwz@jlu.edu.cn).}
\thanks{S. Ma is with the Department of Electrical and Computer Engineering, University of Macau, Macau 999078, China (e-mail: shaodanma@umac.mo).}

}

\maketitle

\begin{abstract}
The optimal design of the energy-efficient multiple-input multiple-output (MIMO) aided uplink ultra-reliable low-latency communications (URLLC) system is an important but unsolved problem.
For such a system, we propose a novel absorbing-Markov-chain-based analysis framework to shed light on the puzzling relationship between the delay and reliability, as well as to quantify the system energy efficiency.
We derive the transition probabilities of the absorbing Markov chain considering the Rayleigh fading, the channel estimation error, the zero-forcing multi-user-detection (ZF-MUD), the grant-free access, the ACK-enabled retransmissions within the delay bound and the interactions among these technical ingredients.
Then, the delay-constrained reliability and the system energy efficiency are derived based on the absorbing Markov chain formulated.
Finally, we study the optimal number of user equipments (UEs) and the optimal number of receiving antennas that maximize the system energy efficiency, while satisfying the reliability and latency requirements of URLLC simultaneously.
Simulation results demonstrate  the accuracy of our theoretical analysis and the effectiveness of massive MIMO in supporting large-scale URLLC systems.
\end{abstract}

\begin{IEEEkeywords}
Ultra-reliable low-latency communications (URLLC), grant-free access, massive MIMO, energy efficiency, absorbing Markov chain.
\end{IEEEkeywords}

%
\IEEEpeerreviewmaketitle

\section{Introduction}
The fifth generation (5G) New Radio (NR) networks are expected to support novel use cases such as tactile internet, factory automation, intelligent transport system and so forth, by relying on the so-called ultra-reliable low-latency communications (URLLC) \cite{URLLC-Bennis}.
Different from usual communication traffic, the URLLC traffic is featured with small packet size of dozens of bytes \cite{URLLC-Power}.
Additionally, URLLC requires to ensure the reliability of at least 99.999\% within the delay bound of 1 ms \cite{3GPP-38913}.
In the uplink of currently deployed LTE networks, the scheduling based medium access schemes require the resource request-and-allocation messaging, which would take at least 10 ms \cite{URLLC-LTE-VTC-2017}.
In addition, the packet size of URLLC traffic may be the same as that of the resource request-and-allocation messages, which deteriorates the spectral efficiency.
Therefore, the scheduling based medium access is not a good solution for achieving uplink URLLC.
\par
To solve this problem, the third generation partnership project (3GPP) proposed a framework of grant-free access, which is a non-scheduling approach without invoking the dynamic and explicit scheduling grant from the base station (BS)~\cite{3GPP-38802}.
In the grant-free access, multiple user equipments (UEs) may send their data packets over the same time-frequency resource block. As a result, collisions may occur, which degrades the reliability performance.
For mitigating this issue, some researchers investigated the benefit of retransmissions to the reliability of grant-free access or random access~\cite{grant-free-VTC, grant-free-ACCESS}.
In \cite{grant-free-VTC}, the authors derived the closed-form expression for the probability of successful transmission subject to the constraint of the maximum number of retransmissions in random access.
Considering Rayleigh fading channels, the authors of \cite{grant-free-ACCESS} studied the reliability performance of two different uplink grant-free schemes, which are based on the stop-and-wait protocol and the diversity transmission approach (i.e., blind retransmission without ACK), respectively.
The authors of \cite{grant-free-VTC} assumed that the reliability was only related to the maximum number of retransmissions, but did not consider the delay bound.
In \cite{grant-free-ACCESS}, the delay bound was mapped into the maximum number of retransmissions, and then the reliability was analysed under the fixed number of retransmissions.
For a general grant-free access system, however, the number of retransmissions is random within the delay bound, which poses challenges on the analysis of the reliability.

\par
In the grant-free access, the reliability can also be improved by reducing the collisions with the multiple-packets-reception (MPR) techniques, which can be realized by multiple-input multiple-output (MIMO) based multi-user detection (MUD) techniques \cite{Shaoshi2015}, such as the zero-forcing (ZF) detector \cite{Linlin-Aloha-WC} and the successive interference cancellation (SIC) detector~\cite{URLLC-2017-GLOBECOM-grant-free}.
Considering the uplink of MIMO systems, the authors of \cite{ URLLC-Pilot} derived the lower bounds of the achievable rates for short packet transmissions using either the maximum-ratio combining (MRC) or the ZF receiver. In addition, how the number of receiving antennas impacts the reliability (or error probability) has also been investigated in \cite{URLLC-MIMO-TVT-2019, URLLC-MIMO-IOT-2020, URLLC-MIMO-IOT-RA-2019, URLLC-2018-CBA, URLLC-MIMO-RE-2019, Linlin-urllc-CL} by different metrics, including the block error rate (BLER) \cite{URLLC-MIMO-TVT-2019, URLLC-MIMO-IOT-2020}, the outage probability \cite{URLLC-MIMO-IOT-RA-2019, URLLC-MIMO-RE-2019}, the collision probability \cite{URLLC-2018-CBA} and the successful transmission probability of a data packet within the delay bound \cite{Linlin-urllc-CL}.
Using either of the above metrics, it has been shown that the reliability can be improved by increasing the number of receiving antennas.

\begin{table}[!t]
	\centering
	\caption{The factors considered in the reliability analysis of MPR-aided grant-free access}
	\renewcommand{\arraystretch}{1.05}
	\label{table-factors-relibility}
	\begin{tabular}{|l|c|c|c|c|c|c|c|c|c|}
		\hline
		& \cite{URLLC-MIMO-TVT-2019}& \cite{URLLC-MIMO-IOT-2020}& \cite{URLLC-MIMO-IOT-RA-2019}& \cite{URLLC-2018-CBA}& \cite{URLLC-MIMO-RE-2019}& \cite{Linlin-urllc-CL} & \tabincell{c}{This\\ paper}\\
		\hline
		Wireless channel  & $\surd$ & $\surd$ & $\surd$ &   & $\surd$ &   & $\surd$\\ \hline
		MUD technique  & $\surd$ & $\surd$ &   & $\surd$ & $\surd$ &   & $\surd$\\  \hline
		Random arrivals   &   &   & $\surd$ & $\surd$ &   & $\surd$ & $\surd$ \\  \hline
		Retransmissions &   &   &   & $\surd$ & $\surd$ & $\surd$ & $\surd$ \\  \hline
		Delay constraints  &   &   &   &   &   & $\surd$ & $\surd$ \\
		\hline
	\end{tabular}
\end{table}

\par
Since the reliability is sensitive to many factors in the MPR-aided system, such as the wireless channel, the MUD technique invoked and so forth, these factors should all be considered in the theoretical analysis, which, however, is an intractable problem.
In Table \ref{table-factors-relibility} we summarize the factors that are considered by the existing work and this paper in the reliability analysis of MPR-aided grant-free access. More specifically, the authors of \cite{URLLC-MIMO-TVT-2019, URLLC-MIMO-IOT-2020} focused on the random feature of wireless channel, but neglected the random feature of packet arrivals and grant-free transmissions, as well as the benefit of retransmissions to reliability.
In \cite{URLLC-MIMO-IOT-RA-2019}, the authors derived the success probability (i.e., $1-$outage probability) of the grant-free random access without retransmissions.
The diversity transmission scheme was considered in~\cite{URLLC-2018-CBA}~\cite{URLLC-MIMO-RE-2019}. In particular, \cite{URLLC-2018-CBA} studied the number of resource blocks needed to support URLLC for the given number of UEs and the average arrival rate.
The authors of \cite{Linlin-urllc-CL} designed an access-probability-adaptive approach for MPR-aided grant-free access, and studied how the reliability is affected by the MPR capability, the number of UEs and the delay bound.
Their results showed that the reliability was improved by the access-probability-adaptive grant-free access scheme compared with the scheme of fixed access probability, and the error probability (i.e., $1-$reliability) roughly had a negative exponential dependence relationship with the delay bound.
Unfortunately, the delay-constrained reliability was overestimated, since the realistic wireless channel characteristics and the specific MIMO-MUD technique invoked to realize MPR were neglected.
It is challenging to analyse the reliability within the delay bound due to multiple stochastic features of the system, such as the random grant-free access, the random number of retransmissions within the delay bound, the sporadic URLLC traffic arrival and the random nature of wireless channels.
Furthermore, the analysis becomes more complicated due to the entanglement between the performance of the grant-free access and that of the MIMO-MUD techniques, as well as owing to the time-dependent feature of the access-probability-adaptive grant-free transmissions.

\par
In addition to satisfying the reliability and latency requirements, designing energy-efficient grant-free access schemes is another challenge for URLLC. Energy-efficient power control and grant-free access schemes for achieving URLLC have been investigated in single-input single-output (SISO) systems~\cite{EE-WCNC-2018, Linlin-martingales-URLLC, URLLC-EE-ICCC2019}. In the context of MIMO systems, the authors of \cite{URLLC-EE-DU} formulated an energy-efficiency maximization problem for URLLC to study how to optimize the number of active antennas, bandwidth allocation, and power control under non-convex quality-of-service (QoS) constraints. The medium access mechanism considered in \cite{URLLC-EE-DU} was grant-based. As far as we know, the energy efficiency optimization problem for URLLC remains unexploited in the MIMO-aided grant-free access system.

\par
In this paper, we focus on the uplink MIMO-aided grant-free access system with URLLC traffic, and devote to providing insights on how the delay-constrained reliability and the system energy efficiency are affected by the system parameters, such as the number of receiving antennas in BS and the number of UEs.
To handle this issue, we propose an absorbing-Markov-chain-based analysis framework, where the grant-free access, the ZF-MUD technique, the $K$-repetition scheme, the ACK-enabled retransmission within the delay bound, the random arrivals of URLLC traffic, the random nature of the wireless channel and the channel estimation error are all considered.
Additionally, the feature of the short packet transmission for URLLC traffic is also taken into account.
Furthermore, we focus on the type of access-probability-adaptive grant-free access scheme, in which the access probability is adjusted according to the number of residual contention UEs and the number of receiving antennas for improving the reliability.
Our main contributions are summarized as follows.
\begin{itemize}
\item We propose an absorbing-Markov-chain-based analytical framework for the ZF-MUD-aided grant-free access system. The entanglement between the ZF-MUD technique and the grant-free access, the multiple stochastic features of the system and their temporal correlations are all characterized in the derivation of the transition probability matrix of the absorbing Markov chain formulated.
\item We solve the challenging delay-constrained reliability and energy-efficiency analysis  problem for the MIMO-aided grant-free access system that supports uplink URLLC traffic, upon using the above absorbing-Markov-chain based analytical framework and transforming the delay constraint into the restrictions imposed on the allowable transition times of the Markov chain.
Simulation results demonstrate that our theoretical analysis of reliability is accurate when the arrival probability is larger than the ratio of the number of receiving antennas to the number of UEs.
\item We present the optimal design of the energy-efficient ZF-MUD-aided grant-free access system with uplink URLLC traffic by solving two energy efficiency maximization problems subject to the QoS constraints of URLLC. Both the optimal number of UEs and that of receiving antennas are found, while fixing the other system parameters. Our simulation results demonstrate that the optimal number of UEs increases almost
linearly with the number of receiving antennas and that massive MIMO constitutes an energy-efficient solution for supporting uplink URLLC in large-scale-antenna based communication systems.
\end{itemize}

\par
The remainder of this paper is organized as follows.
In Section \ref{sec-system}, we present the system model with the ZF-MUD algorithm and the $K$-repetition scheme.
In Section \ref{sec-Markov-design}, we derive the delay-constrained reliability and the energy efficiency based on the absorbing Markov chain theory.
The optimal design of the energy-efficient MIMO-MUD-aided uplink URLLC system is proposed in Section \ref{sec-Markov-design}. Then in Section \ref{sec-simulation}, simulation results are presented to verify our theoretical analysis, and the maximum energy efficiency of the system is also characterized.
Finally, the concluding remarks are offered in Section \ref{sec-conclusion}.

\par
\textit{Notations}: $\textbf{I}_{M}$ represents the $M \times M$ identity matrix. $(\cdot)^{\rm{T}}$, ${(\cdot)}^*$ and $(\cdot)^{\rm{H}}$ denote the operators of the transpose, the conjugate, and the complex conjugate transpose respectively. $\lfloor \cdot \rfloor$ denotes the operator of rounding down. $e$ represents the Euler's number. $[\bm{Z}]_{i,j}$ denotes the element of the matrix $\bm{Z}$ in the $i$-th row and the $j$-th column, and $\left[\textbf{z}\right]_{i}$ denotes the $i$-th element in the vector $\textbf{z}$. In addition, $\{ x \}_+ = \max \{ x, 0 \}$, and ${\parallel{\cdot}\parallel}_1$ denotes the 1-norm.

\section{System model}
\label{sec-system}
In this paper, we focus on the uplink transmission of a single cell composed of one BS and $N$ UEs with URLLC traffic.
The UE is equipped with one transmitting antenna, and the BS is equipped with $M>1$ receiving antennas. Some UEs may have URLLC traffic to transmit simultaneously, and each of these UEs may occupy the same frequency band of $B$~Hz that is  reserved as a resource pool for these $N$ URLLC UEs to contend for.

\par
In 5G NR, the time is divided into a number of subframes, and each of them has a fixed length of $1$\,ms. Furthermore, each subframe can be divided into multiple slots, and the length of a single slot can vary according to the delay requirement of the URLLC traffic and the 5G NR numerology \cite{3GPP-38802}. Typically, each slot consists of multiple short transmission time intervals (S-TTIs). The length of a single S-TTI is denoted by $\tau$, which can be configured to a reduced duration, such as $2$, $3$, or $7$ orthogonal frequency division multiplexing (OFDM) symbols for URLLC \cite{URLLC-2017-ICC}.

\par
We consider the discrete-time event-driven URLLC traffic, which is generated periodically on the UE side, but needs to be transmitted only when a physical event is detected \cite{Linlin-urllc-CL}.
Take the URLLC traffic in tactile internet as an example. The tactile information is perceivable periodically but needs to be transmitted only when the relative difference between two successive stimuli exceeds the just-noticeable difference (JND)~\cite{TI_PROCEEDING_2019}.
Since the arrivals of the URLLC traffic are sporadic, the period between two contiguous arrivals of the traffic is almost surely larger than the delay bound~\footnote{The average arrival rate of sporadic traffic is small, which means that the average inter-arrival time is long. In addition, the delay bound of URLLC traffic is very tight and usually smaller than $1$ ms. Thus, the inter-arrival time is almost surely larger than the delay bound.}.
Assume that the system is tailored to the specific type of URLLC use cases so that the length of a slot equals the delay bound defined as $D^{\max}$.
For each UE, the packet arrival in each slot is assumed to follow the Bernoulli distribution with the parameter $\mu$, which also represents the arrival probability of the URLLC traffic for each UE.

\par
\begin{figure}[!t]
\centering
\includegraphics[width=2.7 in]{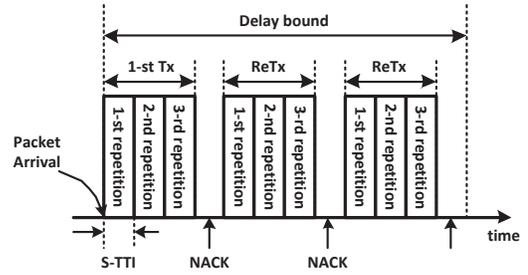}
\caption{An example of $K$-repetition grant-free transmission with $K_{\text{rep}}=3$, $K_{\text{F}}=1$ and the access probability that is equal to $1$.}
\label{fig-ue-Krep}
\end{figure}

In this paper, the UEs that have packets in the buffer are called the contention UEs.
In a S-TTI, the contention UE transmits or retransmits a short packet of $\beta$ bits with a probability, which is called access probability.
In the access-probability-adaptive grant-free access scheme, the access probability is time-varying, and becomes larger when the number of contention UEs decreases or the number of receiving antennas increases.
Furthermore, the $K$-repetition scheme is considered, where the contention UE can be configured to autonomously transmit the same packet for $K_{\text{rep}}$ repetitions in consecutive $K_{\text{rep}}$ S-TTIs, as shown in Fig. \ref{fig-ue-Krep}. It is also noted that each repetition occupies one S-TTI.
The BS utilizes the ZF-MUD algorithm to decode multiple data streams from different UEs, and then utilizes the maximal ratio combining (MRC) to combine the post-processing repetitions of a UE.
After processing the received packets, the BS sends the ACK or NACK feedback so as to inform the UEs if their transmissions are successful.
Similar to \cite{grant-free-ACCESS}, a UE shall wait for the ACK/NACK for a time period that may be larger than or equal to the sum of the processing time at the BS and the ACK/NACK transmission duration. In this paper, the duration of the waiting time is assumed to be $K_\text{F}$ S-TTIs (i.e., $K_\text{F} \tau$).
A UE will stop contention if it receives ACK.
Otherwise, the failed UE will try to retransmit the packet within this slot.
Thus, the number of contention UEs may decline along the S-TTIs within a slot.
Consequently, the access probability may increase along the S-TTIs within this slot.
However, the number of contention UEs may increase or decline across multiple slots.
Considering the time for ACK/NACK feedback, we assume that a slot consists of $L$ S-TTIs, and $L \tau =D^{\max}$. Let $K$ represent the maximum number of transmissions for a packet, and we have $K = \lfloor L/(K_{\text{rep}}+K_{\text{F}}) \rfloor$.

\par
Here, $N'$ UEs out of all the contention UEs are assumed to transmit simultaneously. In this paper, both the path-loss and the small-scale fading are considered.
For the UE $n'$, the path-loss coefficient is defined as $\alpha_{n'}$, and the small-scale fading coefficient vector is defined as $\bm{h}_{n'}$.
The path-loss model (i.e., $10 \lg(\alpha_{n'})$) is assumed to be $-35.3-37.6 \lg(d_{n'})$ [dB], where $d_{n'}$ denotes the distance (in meter) from UE ${n'}$ to the BS.
The small-scale fading is assumed to be Rayleigh fading, hence the elements of $\bm{h}_{n'}$ are independent and identically distributed (IID) complex Gaussian random variables with zero mean and unit variance, i.e., $\bm{h}_{n'} \sim \mathcal{CN}(0, \textbf{I}_{M}) \in \mathbb{C}^{M \times 1}$.
At a given time instant, the received signal vector $\bm{r}$ is given by
\begin{equation}
\label{eq-receive-signal-zf}
\bm{r} = \sum\limits_{n'=1}^{N'} \sqrt{P_{n'}^{\rm{UE}}\alpha_{n'}} \bm{h}_{n'} x_{n'} + \bm{v},
\end{equation}
where $x_{n'}$ denotes the transmitted data symbol of UE ${n'}$, and $\mathbb{E}\left\{x_{n'}^* x_{n'}\right\}=1$. $P_{n'}^{\text{UE}}$ denotes the transmit power of UE ${n'}$, and $\bm{v}$ denotes the additive white noise vector, which follows $\mathcal{CN}(0, N_0 B \textbf{I}_{M})$, with $N_0$ denoting the noise power spectral density.
\par
Considering the features of grant-free short packet transmissions, we assume that the fractional open-loop power control scheme of LTE and NR is used. This scheme only considers the path-loss of the wireless channel, and is also named the full path-loss inversion power control scheme \cite{open-loop-power-control}. In the scheme, the UE compensates for its own path-loss to keep the average received signal power equal to the same threshold $\xi$, i.e.,
\begin{equation}
\label{eq-xi}
P_{n'}^{\rm{UE}}\alpha_{n'} = \xi.
\end{equation}
For clarity, we call $\xi$ the equivalent transmit power (ETP). Then, \eqref{eq-receive-signal-zf} is rewritten as
\begin{equation}
\label{eq-receive-signal-zf-1}
\bm{r} = \sum\limits_{n'=1}^{N'} \sqrt{\xi} \bm{h}_{n'} x_{n'} + \bm{v}= \sqrt{\xi} \bm{H} \bm{x} + \bm{v},
\end{equation}
where $\bm{H}=[\bm{h}_{1}, \cdots, \bm{h}_{N'}]$, and $\bm{x}=[x_{1}, \cdots, x_{N'}]^{\rm{T}}$.

\par
In this paper, the ZF detector is particularly considered for demonstrating the effectiveness of our scheme, since it is widely recognized as the primary detector in the uplink of 5G systems. This is because the ZF detector has a low complexity, yet in particular it can achieve a near-optimal performance when the massive MIMO technology is deployed~\cite{Shaoshi2015}. 
Define $\bm{W}$ as the ZF detection matrix based on the estimated channel $\hat{\bm{H}}$, and $\bm{W}=(\hat{\bm{H}}^{\rm{H}}\hat{\bm{H}})^{-1}{\hat{\bm{H}}}^{\rm{H}}$. In practice, $\hat{\bm{H}}$ can be estimated by the BS via the pilot-assisted channel estimation scheme, and the estimation error is hard to avoid. Similar to \cite{ZF-channel-error}, we model the non-perfect estimated channel as
\begin{equation}
\label{eq-channel-error}
\hat{\bm{H}} = \bm{H} + \delta \bm{\Omega},
\end{equation}
where $\delta \bm{\Omega}$ represents the estimation error that is uncorrelated with $\bm{H}$. The elements of $\bm{\Omega}$ are IID zero-mean complex Gaussian random variables with unity variance; $\delta$ quantifies how accurate the channel estimation is.
The processed signal after using the ZF detector is given as
\begin{equation}
\label{eq-post-processing-signal}
\begin{aligned}
\bm{r}_{\text{ZF}} &= \bm{W} \bm{r} = \bm{W} (\sqrt{\xi} \bm{H} \bm{x} + \bm{v}) \\
&= \sqrt{\xi}\bm{W} \bm{H} \bm{x} + \bm{W} \bm{v}\\
&= \sqrt{\xi}\bm{W} (\hat{\bm{H}}-\delta \bm{\Omega}) \bm{x} + \bm{W} \bm{v}\\
&= \sqrt{\xi}\bm{W}\hat{\bm{H}}\bm{x} + \bm{W} (\bm{v}-\sqrt{\xi}\delta \bm{\Omega} \bm{x})\\
&= \sqrt{\xi} \bm{x} + (\hat{\bm{H}}^{\rm{H}}\hat{\bm{H}})^{-1}{\hat{\bm{H}}}^{\rm{H}} (\bm{v}-\sqrt{\xi}\delta \bm{\Omega} \bm{x}).
\end{aligned}
\end{equation}
Then, the ZF post-processing SNR of the $l$-th repetition of UE ${n'}$ in the case that $N'$ UEs simultaneously transmit is given by \eqref{eq-zf-sinr}.
\begin{figure*}[!t]
\begin{equation}
	\label{eq-zf-sinr}
	\begin{aligned}
		\gamma_l(n'|N') 
		&= \frac{\xi}{\mathbb{E}\left[ \left\| \left[(\hat{\bm{H}}^{\rm{H}}\hat{\bm{H}})^{-1}{\hat{\bm{H}}}^{\rm{H}} (\bm{v}-\sqrt{\xi}\delta \bm{\Omega} \bm{x})\right]_{n',1} \right\|^2 \right] }\\
		&= {\small \frac{\xi}{\mathbb{E}\left[ \left[ (\hat{\bm{H}}^{\rm{H}}\hat{\bm{H}})^{-1}{\hat{\bm{H}}}^{\rm{H}} (\bm{v}-\sqrt{\xi}\delta \bm{\Omega} \bm{x}) ({\bm{v}}^{\rm{H}}-\sqrt{\xi}\delta {\bm{x}}^{\rm{H}} {\bm{\Omega}}^{\rm{H}}) \hat{\bm{H}} (\hat{\bm{H}}^{\rm{H}}\hat{\bm{H}})^{-1}  \right]_{n',n'} \right]} }\\
		&= \frac{\xi}{(N_0 B+\xi \delta^2 N') \left[ (\hat{\bm{H}}^{\rm{H}}\hat{\bm{H}})^{-1} \right]_{n',n'}}.
	\end{aligned}
\end{equation}
	\hrulefill
\end{figure*}
By jointly considering the $K$-repetition scheme, the MRC\&ZF post-processing SNR of the UE ${n'}$ in the case that $N'$ UEs simultaneously transmit is given by
\begin{equation}
\label{eq-mrc-sinr}
\gamma(n'|N') = \sum\limits_{l=1}^{K_{\text{rep}}} \gamma_l(n'|N').
\end{equation}

\par
For the URLLC traffic, since only a small number of bits are transmitted in each coded block, the error probability caused by the channel noise and interference is high due to the use of finite block-length channel codes \cite{URLLC-proceeding}.
Therefore, the analysis of the achievable PHY transmission rate for short packets is fundamentally different from the traditional PHY rate analysis that is based on the classic Shannon channel capacity assuming infinite coded block-length.
Using the normal approximation, the maximal achievable PHY transmission rate in the finite block-length regime is derived by \cite{channel-coding} for SISO systems with transmit power constraints and isotropically distributed codebooks. This result is also suitable for the single-input-multiple-output (SIMO) and multiple-input single-output (MISO) systems \cite{URLLC-magzine}.
For the given BLER $\varepsilon^{\rm{B}}$ and the given SNR $\gamma(n'|N')$, the maximal achievable rate [bits/s] of UE $n'$ with finite block-length can be approximated as \cite{channel-coding, URLLC-magzine}:
\begin{equation}
\label{eq-rate-bits}
\begin{aligned}
	& R(\gamma(n'|N')) \approx \\
	& B {\displaystyle \left\{ {{{\log }_2}\left( {1 + {\gamma(n'|N')}} \right) - \sqrt {\frac{V(\gamma(n'|N'))}{\tau B}\left( {{{\log }_2}e } \right)}f_{\rm{Q}}^{ - 1}(\varepsilon^{\rm{B}}) } \right\}_ + },
\end{aligned}
\end{equation}
where $f_{\rm{Q}}^{ - 1}(x)$ denotes the inverse of the Gaussian-Q function, and $V(\gamma(n'|N'))$ denotes the channel dispersion, which characterizes the random variability of a channel with respect to a deterministic channel with the same capacity \cite{URLLC-Secure}. $V(\gamma(n'|N'))$ is expressed as
\[
\begin{aligned}
	\displaystyle V(\gamma(n'|N')) &= \frac{(2 + {\gamma(n'|N')}){\gamma(n'|N')}}{{(1 + {\gamma(n'|N')})}^2} \\
	&= 1-\frac{1}{(1+\gamma(n'|N'))^2}.
\end{aligned}
\]
For the reader's convenience, we restate the definitions of $B$ and $\tau$ here. $B$ denotes the bandwidth dedicated to the uplink URLLC traffic, and $\tau$ denotes the duration of one S-TTI.
\par
Please note that $N'$ is a random variable because of the sporadic arrivals of the URLLC traffic and the random nature of the access-probability-adaptive grant-free transmission.
A transmission is regarded as successful when the achievable PHY rate is equal to or larger than the required one (i.e., $R(\gamma(n'|N')) \geq \beta/\tau$)~\footnote{According to the finite block-length information theory\cite{channel-coding}, if the achievable transmission rate is larger than or equal to the target $\beta/\tau$, a modulation and coding scheme can be found to transmit this packet in the S-TTI with the BLER smaller than $\varepsilon^{\rm{B}}$, which is regarded as a successful transmission in this paper.}.
If a packet is not transmitted successfully within the slot, the packet will be dropped.
From \cite{3GPP-38913}, the reliability can be evaluated by the probability of successfully transmitting a packet within a certain delay bound. In this paper, the reliability of the system is evaluated by a stricter metric, which is the probability of all packets transmitted successfully within a slot.
Since the reliability and the delay are intertwined, herein we use the term of delay-constrained reliability.

\section{Performance analysis and optimization of the ZF-MUD-aided grant-free access system}
\label{sec-Markov-design}
In this section, we first present the absorbing-Markov-chain-based performance analysis framework, which helps us derive the delay-constrained reliability and system energy efficiency for the grant-free access that invokes the ZF-MUD algorithm and the $K$-repetition scheme. Then based on these results, the system energy efficiency is further optimized.

\subsection{Analysis of the delay-constrained reliability}

Let $\varepsilon$ denote the error probability within the delay bound. Then $1-\varepsilon$ is the delay-constrained reliability of the system. Define $p_n$ as the probability of $n$ packets arriving in the first S-TTI of a slot, and define $\lambda_n$ as the conditional probability of all the packets successfully transmitted within the slot in the case that $n$ packets arrive in the slot. For the given length of the slot, we have
\begin{equation}
\label{relibility-ini}
\varepsilon = \sum\limits_{n=1}^N p_n \left(1-\lambda_n\right) .
\end{equation}
For the convenience of readers, we restated the definition of $N$, which denotes the number of UEs in the system.

\par
In this paper, the packet arrival of UE $n'$ in each slot is assumed to follow the Bernoulli distribution with the parameter $\mu_{n'}$. Let ${\bm{\sigma}} \in { \{0, 1\} } ^N$ represent the vector of the arrival indicators. If the UE $n'$ has a new packet arrival, we have $\sigma_{n'} = 1 $. Otherwise, we have $\sigma_{n'} = 0 $. Additionally, we assume that the arrivals of packets are independent among UEs. Hence, $p_n$ is given as
\begin{equation}
	\label{eq-pr-n-0}
	p_n=\sum\limits_{{\parallel{\bm{\sigma}}\parallel}_1=n} \prod\limits_{{n'}=1}^N [(1-\sigma_{n'})(1-\mu_{n'}) + \sigma_{n'} \mu_{n'}],
\end{equation}
and when $\mu_{n'}=\mu$, we have
\begin{equation}
\label{eq-pr-n-transmiting}
p_n = \begin{pmatrix}
N\\
n
\end{pmatrix} \mu^n (1-\mu)^{N-n},\;\; 0 < n \leq N.
\end{equation}
$p_n$ is also equal to the probability of $n$ contention UEs existing in the first S-TTI of a slot.
\par
Next, let us move on to the challenging job of deriving the conditional probability $\lambda_n$ that all the packets are  transmitted successfully within the slot, when there are $n$ packets arriving in the slot. By considering the multiple stochastic features in the ZF-MUD-aided grant-free access with the $K$-repetition scheme, we find that $\lambda_n$ is influenced by the following factors.
\begin{itemize}
\item The wireless channel is random, which may result in failure of the transmission of a packet.
\item The number of simultaneously transmitting UEs in a S-TTI is random, and its statistical property is determined by the random grant-free access and the number of residual contention UEs in the S-TTI.
\item In the ACK-enabled retransmissions, the number of contending UEs is random and time-dependent along S-TTIs, which also results in the time-varying feature of the access probability for the access-probability-adaptive grant-free access.
\item The allowable maximum number of transmissions for a packet is determined by both the delay bound and the number of repetitions.
\end{itemize}
Therefore, a foundational task for deriving $\lambda_n$ is to characterize the statistical property of the number of contention UEs within a slot.
\par
\begin{figure}[!t]
\centering
\includegraphics[width=2.6 in]{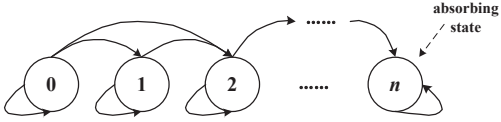}
\caption{An illustration of the absorbing Markov chain $\{X_n(k)\}$.}
\label{fig-absorbing-Markov}
\end{figure}

The absorbing Markov chain theory applies to the statistical analysis in the bounded time.
An absorbing Markov chain is a Markov chain that has at least one absorbing state. An absorbing state is a state that, once entered, it is impossible to leave.
Considering the time-dependent feature, we utilize the absorbing Markov chain $\{X_n(k)\}$ to model the evolution of the total number of UEs that transmit successfully in the S-TTI $k(K_{\text{rep}}+K_{\text{F}})$ for ($0 \leq k \leq K$)\footnote{The S-TTI $0$ represents the beginning of the first S-TTI in the slot.}, under the condition that $n$ contention UEs exist at the beginning of the slot. In this paper, we focus on the state-jump of $\{X_n(k)\}$ after each transmission, which consists of $K_{\text{rep}}$ repetitions. Thus, the states of $\{X_n(k)\}$ represent the accumulated number of UEs transmitting successfully in the S-TTI $k(K_{\text{rep}}+K_{\text{F}})$, $X_n(k) \in \{0, 1, \cdots, n\}$ and $X_n(0)=0$. As a result, the number of contention UEs is $n-X_n(k)$ from the S-TTI $(k(K_{\text{rep}}+K_{\text{F}})+1)$ to the S-TTI $(k+1)(K_{\text{rep}}+K_{\text{F}})$.
The absorbing state of $\{X_n(k)\}$ is the state $n$, because at most $n$ UEs transmit successfully under the condition that $n$ contention UEs exist at the beginning of the slot.
For an absorbing Markov chain, it is possible to go from any state to an absorbing state within multiple jumps\footnote{Here, ``jump'' means that the transition from one state to another.}. However, the absorbing Markov chain may not enter into the absorbing state if the number of jumps is constrained.
Due to the latency bound, the allowable maximum transition times of the absorbing Markov chain $\{X_n(k)\}$ is $K$, which is equal to $\lfloor L/(K_{\text{rep}}+K_{\text{F}}) \rfloor$.
The Markov chain $\{X_n(k)\}$ may not enter into the absorbing state $n$ within $K$ jumps, which means that some packets are dropped at the end of the slot.
The formulated absorbing Markov chain is shown in Fig. \ref{fig-absorbing-Markov}.
\par
Let $\textbf{P}_n$ denote the transition matrix of our absorbing Markov chain in the case that $n$ contention UEs exist at the beginning of a slot. The element of $\textbf{P}_n$ in the $(i+1)$-th row and the $(j+1)$-th column represents the probability that the total number of successfully transmitting UEs changes from $i$ in the $k(K_{\text{rep}}+K_{\text{F}})$-th S-TTI to $j$ in the $(k+1)(K_{\text{rep}}+K_{\text{F}})$-th S-TTI, where we have $0 \leq i \leq n$, $0 \leq j \leq n$, and $0 \leq k \leq K-1$.
In this paper, the transmission of a packet for a UE is regarded as successful if its achievable transmission rate is not smaller than the threshold $\beta/\tau$. 
Thus, the transition probability $\Pr \left\{ {{X_n(k + 1)} = j|X_n(k) = i} \right\}$ in $\textbf{P}_n$ is affected by the ZF-MUD, the $K$-repetition scheme, the specific grant-free access scheme and their interaction.
To derive $\Pr \left\{ {{X_n(k + 1)} = j|X_n(k) = i} \right\}$, three cases are discussed as follows.
\par
\textit{(1) Case of $i < j \leq \min\{n, i+M\}$}: This case means that $j-i$ UEs transmitting successfully when there have been $i$ contention UEs transmitting successfully till the $k(K_{\text{rep}}+K_{\text{F}})$-th S-TTI. Then, the total number of UEs transmitting successfully is equal to $j$ in the $(k+1)(K_{\text{rep}}+K_{\text{F}})$-th S-TTI.
With consideration of the ZF-MUD and the computation complexity, the number of UEs transmitting successfully is regarded as not larger than $M$. Hence, we have $j \leq \min\{n, i+M\}$.
Let $p^{\text{TX}}(N'|n-i)$ denote the probability of the event that $N'$ out of $n-i$ residual contention UEs transmit.
Let $p^{\text{SUCC}}(j-i|N')$ denote the probability of the event that $j-i$ contention UEs transmit successfully under the condition that $N'$ contention UEs transmit.
In this case, the transition probability $\Pr \left\{ {{X_n(k + 1)} = j|X_n(k) = i} \right\}$ is given as
\begin{equation}
	\label{eq-transition-pr-0}
		\begin{array}{l}
			\Pr \left\{ {{X_n(k + 1)} = j|X_n(k) = i} \right\} =\\
			\left\{
			\begin{array}{l}
				\displaystyle \sum\limits_{N'=j-i}^{\min\{n-i, M\}} p^{\text{TX}}(N'|n-i) p^{\text{SUCC}}(j-i|N'), f_{\rm{p}}(n-i)<1,\\
				\displaystyle p^{\text{SUCC}}(j-i|n-i), f_{\rm{p}}(n-i)=1,
			\end{array}
			\right.
		\end{array}
\end{equation}
where $f_{\rm{p}}(n-i)$ denotes the access probability of UE. $f_{\rm{p}}(n-i)$ is a function with respect to the number of residual contention UEs (i.e., $n-i$) for the given number of receiving antennas. In \cite{MPR-Access-YunH2014}, it has been proved that the success probability of the MPR-aided grant-free transmissions is greatly improved when $f_{\rm{p}}(n-i)$ is set to $\min\{M/(n-i+1),\; 1\}$.
\par
Since each contention UE transmits the repetitions of data packet with a probability $f_{\rm{p}}(n-i)$ independently, the transmission behaviour of the UE can be modelled as a Bernoulli distribution with the parameter $f_{\rm{p}}(n-i)$. Then, the probability $p^{\text{TX}}(N'|n-i)$ is given by
\begin{equation}
\label{eq-transition-pr-tmp1}
p^{\text{TX}}(N'|n-i) = \displaystyle \begin{pmatrix}
n-i\\
N'
\end{pmatrix} {\left(f_{\rm{p}}(n-i) \right)}^{N'}{\left( {1 - f_{\rm{p}}(n-i)} \right)}^{n-i-N'}.
\end{equation}
Define the conditional outage probability of a UE in the case that $N'$ UEs transmit simultaneously as $\rho(N')=\Pr\left\{ \displaystyle R(\gamma(n'|N')) < \beta/\tau  \right\}$.
Then, $p^{\text{SUCC}}(j-i|N')$ equals the probability that $N'-(j-i)$ out of $N'$ UEs suffer outage. Thus, $p^{\text{SUCC}}(j-i|N')$ is given by
\begin{equation}
\label{eq-transition-pr-tmp2}
 p^{\text{SUCC}}(j-i|N') = \displaystyle \begin{pmatrix}
{N'}\\
{j - i}
\end{pmatrix} {(1-\rho(N'))}^{j-i} {\rho(N')}^{N'-(j-i)}.
\end{equation}

\begin{theorem}
\label{lemma-outage}
When invoking the ZF-MUD algorithm and the short packet transmission under the assumption of Rayleigh fading, $\rho(N')$ is given by
\begin{equation}
\label{eq-outage-pr-easy}
\rho(N') = \frac{\displaystyle f_{\gamma}\left((M-N'+1) K_{\text{rep}}, \frac{(N_0 B + \xi \delta^2 N')(2^{\Omega}-1)}{\xi(\delta^2+1)} \right)}{f_{\Gamma}((M-N'+1)K_{\text{rep}})},
\end{equation}
and
\begin{equation}
\Omega=\displaystyle \sqrt {\frac{{{\log }_2}e}{\tau B}}f_{\rm{Q}}^{ - 1}(\varepsilon^{\rm{B}})+\frac{\beta}{\tau B}
\end{equation}
where $f_{\gamma}(\cdot,\cdot)$ is the lower incomplete Gamma function, and $f_{\Gamma}(\cdot)$ is the Gamma function.
\end{theorem}
\begin{IEEEproof}
Please see Appendix \ref{app-outage-probability}.
\end{IEEEproof}

\par
\textit{(2) Case of $j < i$ or $j \geq i + M + 1$}: The case of $j < i$ will not happen because the total number of UEs that transmit successfully does not decrease along the S-TTIs within the slot. 
Additionally, we consider that the case of $j \geq i + M + 1$ will not happen because of the ZF-MUD technique\footnote{The assumption that the case $j-i > M$ will not happen is conservative, due to the benefits from the $K$-repetition scheme for $K_{\text{rep}}>1$.}. 
Thus, we have
\begin{equation}
\label{eq-transition-pr-1}
\Pr \left\{ {X_n(k + 1) = j|X_n(k) = i} \right\} = 0.
\end{equation}

\par
\textit{(3) Case of $j=i$}: This case means that all transmissions of packets fail in the $(k+1)(K_{\text{rep}}+K_{\text{F}})$-th S-TTI under the condition of $i$ UEs transmitting successfully till the $k(K_{\text{rep}}+K_{\text{F}})$-th S-TTI.
In this case, the transition probability is given as
\begin{equation}
\label{eq-transition-pr-2}
\begin{array}{l}
\Pr \left\{ {X_n(k + 1) = i|X_n(k) = i} \right\} = \\
\qquad 1 - \sum\limits_{j = i + 1}^{\min\{n, i+M\}} {\Pr \left\{ {X_n(k + 1) = j|X_n(k) = i} \right\}} .
\end{array}
\end{equation}

\par
According to \cite{Markov}, the canonical form of the matrix $\textbf{P}_n$ is given by
\begin{equation}
\label{eq-absorb-matrix}
\textbf{P}_n = \left( {\begin{array}{*{20}{c}}
\textbf{Q}_n&|&\textbf{y}_n\\
{ -  - }&|&{ -  - }\\
\textbf{0}&|&\textbf{I}_1
\end{array}} \right),
\end{equation}
where $\textbf{Q}_n$ is the transition sub-matrix that only involves the transition probabilities associated with the transient states. $\textbf{y}_n$ is a column vector that involves the transition probabilities from the transient states to the absorbing state. Since there is only one absorbing state here, $\textbf{0}$ is a null vector and $\textbf{I}_1$ is the identity matrix with only one element $1$.
For clarity, we present an example of the transition matrix $\textbf{P}_n$ for $n=3$ in the case of $M=2$, as shown in \eqref{eq-P3}. In this example, $[\textbf{P}_3]_{k,k} = 1- \sum\limits_{m=k+1}^{\min\{k+2,4\}} [\textbf{P}_3]_{k,m}$ for $1 \leq k \leq 3$, and $p^{\text{TX}}(N'|n-i)$ and $p^{\text{SUCC}}(j-i|N')$ can be obtained by \eqref{eq-transition-pr-tmp1} and \eqref{eq-transition-pr-tmp2} respectively.
\begin{figure*}[!t]
\normalsize
\small
\begin{equation}
\label{eq-P3}
 \textbf{P}_3 = \left[ {\begin{array}{*{10}{c}}
   {[\textbf{P}_3]_{1,1}} & {\sum\limits_{N' = 1}^2 {p^{\text{TX}}(N'|3)p^{\text{SUCC}}(1|N')} } & { p^{\text{TX}}(2|3)p^{\text{SUCC}}(2|2)}  & | & 0  \\
   0 & {[\textbf{P}_3]_{2,2}} & {\sum\limits_{N' = 1}^2 {p^{\text{TX}}(N'|2)p^{\text{SUCC}}(1|N')} } & | &  {p^{\text{TX}}(2|2)p^{\text{SUCC}}(2|2)}  \\
   0 & 0 & {[\textbf{P}_3]_{3,3}} & | &  p^{\text{SUCC}}(1|1)  \\
{- - - - - -} & {- - - - - - - - - - - -} &{- - - - - - - - - - - -}&|&{- - - - - - - -}\\
   0 & 0 & 0 & | & 1  \\
\end{array}} \right]
\end{equation}
\hrulefill
\end{figure*}
\par
In this paper, the grant-free transmissions are regarded as unreliable if some packets are dropped at the end of the slot, which means that the absorbing Markov chain $\{X_n(k)\}$ may not get into the absorbing state $n$ within $K$ jumps.
Now let us define the stopping time $T_n$ as the first time instant when the Markov chain with the state space of $\left\{ 0, \; 1, \; \cdots, \; n \right\}$ enters into its absorbing state, i.e., $X_n(k)=n$. Then, $T_n$ is given as
\begin{equation}
\label{eq-stopping-time}
T_n = \min \{k| \;\; X_n(k)=n, \;\; k \geq 0\}.
\end{equation}
Therefore, the conditional probability $\lambda_n$ is equal to the probability of the event that $T_n \leq K$, under the condition that $n$ contention UEs exist at the beginning of a slot. Then the conditional probability $\lambda_n$ is given by
\begin{equation}
\label{eq-lambda-n}
\begin{aligned}
\lambda_n &= \Pr\{T_n \leq K\}
=\sum\limits_{k=1}^K \Pr\{T_n=k\} \\
& \overset{(a)}{=} \sum\limits_{k=1}^{K} \Pr\{X_n(k)=n|X_n(0)=0\} \\
& =\sum\limits_{k=0}^{K-1} \left[ (\textbf{Q}_n)^{k} \textbf{y}_n \right]_{1}.
\end{aligned}
\end{equation}
The event ``$T_n=k$'' means that the absorbing Markov chain $\{X_n(k)\}$ enters into the absorbing state $n$ after $k$ jumps for $1 \leq k \leq K$. Thus, we have $\Pr\{T_n=k\}=\Pr\{X_n(k)=n|X_n(0)=0\}$, and the equality (a) holds. Additionally, $\left[ (\textbf{Q}_n)^{k} \right]_{i+1, j+1}$ represents the probability of hitting the transient state $j$ after $k$ jumps from the transient state $i$. Thus,
$\left[ (\textbf{Q}_n)^{k} \textbf{y}_n \right]_{1}$ represents the probability of the Markov chain $\{X_n(k)\}$ getting into the absorbing state $n$ after $k+1$ jumps (for $0 \leq k \leq K-1$) from the initial state $0$, which also represents the probability of $n$ UEs transmitting successfully till the S-TTI $(k+1)(K_{\text{rep}}+K_{\text{F}})$ for $0 \leq k \leq K-1$.
\par
Substitute (\ref{eq-pr-n-transmiting}) and (\ref{eq-lambda-n}) into (\ref{relibility-ini}), then we have
\begin{equation}
\label{eq-reliability-delay}
\begin{aligned}
	\varepsilon &= \sum\limits_{n=1}^N p_n \left[{1- \sum\limits_{k=0}^{K-1} \left[ (\textbf{Q}_n)^{k} \textbf{y}_n \right]_{1} } \right] \\
	& \overset{(a)}{=}
	\sum\limits_{n=1}^N \begin{pmatrix}
		N\\
		n
	\end{pmatrix} \mu^n (1-\mu)^{N-n} \left[{1- \sum\limits_{k=0}^{K-1} \left[ (\textbf{Q}_n)^{k} \textbf{y}_n \right]_{1} } \right].
\end{aligned}
\end{equation}
The equality (a) holds in the case of $\mu_{n'}=\mu$ for $1 \leq n' \leq~N$.
From \eqref{eq-reliability-delay}, we can see that the delay-constrained error probability of the grant-free system is impacted by the delay bound about $ K(K_{\text{rep}}+K_{\text{F}})\tau$, the arrival probability $\mu$, the noise power $N_0B$, the ETP $\xi$, the packet size $\beta$, the number of channel uses $\tau B$ for the transmission of a packet, the total number of UEs $N$ and the number of receiving antennas $M$.

\begin{remark}
Our analysis framework can also be directly extended to other linear detection algorithms, such as the MRC-MUD and the MMSE-MUD, in addition to the ZF-MUD. In this context, only the conditional outage probability $\rho(N')$ needs to be re-calculated. Let us take the MRC-MUD as an example. The MRC-MUD detection matrix is $\bm{W}=\hat{\bm{H}}^{\rm{H}}$, and the MUD post-processing SNR $\gamma_l(n'|N')=\xi {\parallel {\hat{\bm{h}}}_{n'} \parallel}^4/(\xi \sum\limits_{i=1,i\neq n'}^{N'}{\parallel {\hat{\bm{h}}}_{n'}^{\rm{H}}{\hat{\bm{h}}}_{i} \parallel}^2 +(\xi \delta^2 N'+N_0B){\parallel {\hat{\bm{h}}}_{n'} \parallel}^2)=\xi {\parallel {\hat{\bm{h}}}_{n'} \parallel}^2/(\xi \sum\limits_{i=1,i\neq n'}^{N'} \frac{{\parallel {\hat{\bm{h}}}_{n'}^{\rm{H}}{\hat{\bm{h}}}_{i} \parallel}^2}{{\parallel {\hat{\bm{h}}}_{n'} \parallel}^2} +\xi \delta^2 N'+N_0B)$. The analysis on the probability distribution function (PDF) of $\gamma_l(n'|N')$ is given in \cite{MRC-SINR}, and then the PDF of $\sum\limits_{l=1}^{K_{\text{rep}}} \gamma_l(n'|N')$ can be calculated. Finally, $\rho(N')$ can be obtained.
\par
In principle, our analysis framework can also be extended to nonlinear MUDs, such as SIC detector. However, this extension is more challenging. Let us consider the extension to the SIC detector as an example. In our analysis, the impact of MUD on reliability is mainly reflected by the conditional outage probability, i.e., $\rho(N')$. Here, $N'$ denotes the number of UEs that transmit simultaneously at a given time instant. The key step of analyzing $\rho(N')$ is to derive the PDF of post-processing SNR. In the analysis of the PDF of post-processing SNR, only the randomness of $N'$ is considered for the linear MUD. However, for the SIC
detector, because the performance of SIC highly relies on the detection ordering scheme, not only the randomness of $N'$ but also the specific ordering of the channel vectors of these UEs should be considered. Thus, the reliability analysis of the uplink MIMO system with SIC detector is another challenging problem.
\end{remark}

\subsection{Analysis of the energy efficiency }
\label{sec-sub-EE}
In this paper, the energy efficiency of the system (denoted by $\eta$) is defined as follows:
\begin{equation}
\label{eq-ee}
\eta = \varphi/\Psi,
\end{equation}
where $\varphi$ denotes the throughput of the system, and $\Psi$ denotes the average energy consumption of the system. In terms of the system energy consumption, we mainly consider the sum of transmitting power of UEs, and the power consumed by the receiving antennas and consumed by sending the feedback at the BS.
\par
In this paper, the transmission state of packets within a slot can be divided into three categories: (1) all packets are dropped, (2) some packets are dropped and some packets are transmitted successfully, (3) all packets are transmitted successfully within the delay bound.
In the third category, if $n$ contention UEs exist at the beginning of a slot, the average amount of successfully transmitted information is $n(1 - \varepsilon) \beta$ [bits]. In the second category, we need to analyse the statistical property of the total number of packets transmitted successfully in a slot.
In the first category, the transmissions consume energy but do not contribute to the system throughput. We need to analyse the statistical property of the number of residual contention UEs in each transmission for calculating the energy consumption.
\par
Define $p_{i|n}^{\rm{s}}$ for $1 \leq i \leq n-1$ as the probability of the event that $i$ UEs transmit successfully within a slot under the condition that $n$ contention UEs exist at the beginning of the slot. Then we have
\begin{equation}
\label{eq-pr-i-UEs}
p_{i|n}^{\rm{s}} = \left[ ({\textbf{Q}}_n)^K \right]_{1, i+1},
\end{equation}
where $\left[ ({\textbf{Q}}_n)^K \right]_{1, i+1}$ represents the probability of the event that the absorbing Markov chain $\{X_n(k)\}$ hits the transient state $i$ ($1 \leq i \leq n-1$) after the $K$ times of jumps from the initial state $0$, which means that $i$ packets are transmitted successfully in the slot. In this case, the average amount of successfully transmitted information is $\sum\limits_{i = 1}^{n-1} i p_{i|n}^{\rm{s}} \beta$ [bits].
Then the system throughput $\varphi$ is given by
\begin{equation}
\label{eq-throughput}
\begin{aligned}
	\varphi &= \dfrac{\displaystyle \sum\limits_{n = 1}^N {{p_n}} \left[ {\sum\limits_{i = 1}^{n-1} i p_{i|n}^{\rm{s}} + n(1 - \varepsilon )} \right] \beta}{\displaystyle (K_{\text{rep}}+K_{\text{F}}) K\tau} \\
	&= \dfrac{\displaystyle \sum\limits_{n = 1}^N {{p_n}} \left[ {\sum\limits_{i = 1}^{n-1} i \left[ ({\textbf{Q}}_n)^K \right]_{1, i+1} + n(1 - \varepsilon )} \right] \beta}{\displaystyle (K_{\text{rep}}+K_{\text{F}})K\tau}.
\end{aligned}
\end{equation}
The expressions of $p_n$ and $\varepsilon$ can be found in (\ref{eq-pr-n-transmiting}) and (\ref{eq-reliability-delay}), respectively.
\par
Next, let us calculate the average energy consumption of the system. Suppose that $P^{\text{UE-Sum}}$ denotes the sum of the average transmit power of UEs, $P^{\text{BS-T}}$ denotes the average transmit power of the BS sending ACK/NACK, and $P^{\text{BS-C}}$ denotes the circuit power consumed by the antennas at the BS. Here, the transmission time of the ACK/NACK is assumed to be one S-TTI.
Then, the average system energy consumption $\Psi$ is given as
\begin{equation}
\label{eq-system-energy}
\Psi = \frac{K_{\text{rep}}}{K_{\text{rep}}+K_{\text{F}}} P^{\text{UE-Sum}} + \frac{1}{K_{\text{rep}}+K_{\text{F}}} P^{\text{BS-T}} + P^{\text{BS-C}}.
\end{equation}
Define $P_n^{\text{UE-Sum}}$ as the sum of the average transmitting power of UEs in the case that there are $n$ contention UEs at the beginning of a slot. $P^{\text{UE-Sum}}$ is then given by
\begin{equation}
\label{eq-UEs-energy}
P^{\text{UE-Sum}} = \sum\limits_{n=1}^N p_n P_n^{\text{UE-Sum}}.
\end{equation}
\par
Let $\bar P^{\text{UE}}$ denote the average transmit power of UEs, and we have ${\bar P^{\text{UE}}} =\displaystyle\frac{1}{N}\sum\limits_{n'=1}^N P_{n'}^{\text{UE}}$. 
Then, $P_n^{\text{UE-Sum}}$ equals the product of $\bar P^{\text{UE}}$ and the average number of UEs transmitting simultaneously in a S-TTI. Obviously, in the first S-TTI, the average number of UEs transmitting simultaneously equals $nf_{\rm{p}}(n)$. From the S-TTI $(K_{\text{rep}}+K_{\text{F}})k+1$ to the S-TTI $(K_{\text{rep}}+K_{\text{F}})(k+1)$ for $1 \leq k \leq K-1$, the average number of UEs transmitting simultaneously equals $(n-i)f_{\rm{p}}(n-i)$ if there exist $n-i$ residual contention UEs. And
the probability of having $n-i$ contention UEs is calculated as $\left[ ({\textbf{Q}}_n)^k \right]_{1, i+1}$ for $0 \leq i \leq n-1$. Thus, $P_n^{\text{UE-Sum}}$ is given as
\begin{equation}
\label{eq-UE-power-n}
\begin{aligned}
& P_n^{\text{UE-Sum}} = \\
& \displaystyle \frac{ {\bar P^{\text{UE}}} }{K} \left[ n f_{\rm{p}}(n) + \sum\limits_{k=1}^{K-1}\sum\limits_{i=0}^{n-1} \left[ ({\textbf{Q}}_n)^k \right]_{1, i+1} (n-i) f_{\rm{p}}(n-i) \right].
\end{aligned}
\end{equation}

\par
In each slot, if all the packets of URLLC traffic are transmitted successfully before the end of the slot, the BS can become idle or serve other traffic until the next slot begins.
In this paper, we only consider the energy consumption of the URLLC traffic.
Let $T_n^{\rm{BS}}$ denote the percentage of time when the BS serves the URLLC traffic, under the condition that there are $n$ contention UEs at the beginning of the slot.
Then we have
\begin{equation}
\label{eq-BS-power-ACK}
P^{\text{BS-T}} = P^{\text{BS-tx}} \sum\limits_{n=1}^N p_n T_n^{\rm{BS}},
\end{equation}
and
\begin{equation}
\label{eq-BS-power}
P^{\text{BS-C}} = M P^{\text{ca}} \sum\limits_{n=1}^N p_n T_n^{\rm{BS}},
\end{equation}
where $P^{\text{BS-tx}}$ represents the transmit power of the BS for sending ACK/NACK, and $P^{\text{ca}}$ is the circuit power consumed per antenna at the BS. 
$T_n^{\rm{BS}}$ also represents the average number of transitions used until the absorption of $\{X_n(k)\}$ or the delay bound violation. 
Then, $T_n^{\rm{BS}}$ is given by
\begin{equation}
\label{eq-BS-time-n}
T_n^{\rm{BS}} =\sum\limits_{k=0}^{K-1} (k+1)\left[(\textbf{Q}_n)^{k} \textbf{y}_n \right]_{1} + (1-\lambda_n)K.
\end{equation}
Furthermore, putting the above derived results into \eqref{eq-system-energy}, the average energy consumption of the system is given as \eqref{eq-energy-final}. 
\begin{figure*}[!t]
	\begin{equation}
		\label{eq-energy-final}
		\begin{aligned}
			\Psi & = \frac{ {\bar P^{\text{UE}}} K_{\text{rep}} }{K(K_{\text{rep}}+K_{\text{F}})}   
			\sum\limits_{n=1}^N p_n \displaystyle \left[ n f_{\rm{p}}(n) +  \sum\limits_{k=1}^{K-1}\sum\limits_{i=0}^{n-1} \left[ ({\textbf{Q}}_n)^k \right]_{1, i+1} (n-i) f_{\rm{p}}(n-i) \right] \\
			& + \left(\frac{P^{\text{BS-tx}}}{K_{\text{rep}}+K_{\text{F}}}+MP^{\text{ca}}\right)\sum\limits_{n=1}^N p_n \left[ \sum\limits_{k=0}^{K-1} (k+1)\left[(\textbf{Q}_n)^{k} \textbf{y}_n \right]_{1} + (1-\lambda_n)K \right] .
		\end{aligned}
	\end{equation}
	\hrulefill
\end{figure*}
Then, substitute \eqref{eq-throughput} and \eqref{eq-energy-final} into \eqref{eq-ee}, we can obtain the energy efficiency of the system.
Additionally, the expressions of $p_n$ and $\varepsilon$ still can be found in (\ref{eq-pr-n-transmiting}) and (\ref{eq-reliability-delay}), respectively.

\subsection{Energy efficiency optimization}
Let $\varepsilon ^{\max }$ denote the maximum acceptable error probability within the delay bound. For ensuring the delay-constrained reliability of URLLC, $\varepsilon \leq \varepsilon ^{\max }$ has to hold.
In this subsection, we first find the optimal number of UEs (i.e., $N^*$) that maximizes the energy efficiency of the system under the constraint of $\varepsilon \leq \varepsilon ^{\max }$ when the other parameters are fixed.
In this case, the optimization variable of the optimization problem is the number of UEs $N$. Therefore, we rewrite the delay-constrained error probability $\varepsilon$ as $\varepsilon(N)$, and the system energy efficiency $\eta$ as $\eta(N)$.

\par
Define $N^*$ as the optimal number of UEs that maximizes the energy efficiency of the system with URLLC traffic. 
The optimization problem is formulated as a reliability and delay-QoS-constrained energy efficiency maximization problem, which is given by
\begin{equation}
\label{eq-OP-N}
\renewcommand{\arraystretch}{1.1}
\begin{array}{l}
\mathop {\max }\limits_{N} \;\eta(N)\\
\text{s.t.}\quad \varepsilon(N) \leq \varepsilon ^{\max }, N \in \mathbb{Z}^{+}.\\
\end{array}
\end{equation}
Obviously, $\varepsilon(N)$ is an increasing function with respect to $N$. Thus, we first utilize the dichotomy algorithm to search the maximum number of UEs that can be accommodated under the constraint of $\varepsilon(N) \leq \varepsilon ^{\max }$. This constrained maximum number of UEs is defined as $N^{\max}=\max\{N|\varepsilon(N) \leq \varepsilon ^{\max }, N \in \mathbb{Z}^{+}\}$. The framework of the dichotomy algorithm is shown in Table \ref{table-dichotomy}. Then, the exhaustive search method is utilized to find $N^*$ in the interval from $1$ to $N^{\max}$. The total computational complexity of the search processes including the dichotomy algorithm and the exhaustive search method is $O(\log_2{(N^{\text{upper}}-1)})+O(N^{\max})$. Here, $N^{\text{upper}}$ denotes the initial upper bound of the number of UEs in the dichotomy algorithm, and its value is set empirically.

\begin{table}[!t]
\centering
\caption{The dichotomy algorithm for searching in the feasible region of the optimization problem \eqref{eq-OP-N}}
\renewcommand{\arraystretch}{1.02}
\label{table-dichotomy}
\begin{tabular*}{8.0cm}{l}
\hline
$N^{\max}=0$; \\
$N^{\rm{lower}}=1$; /* lower bound on the number of UEs $N$ */\\
$N^{\rm{upper}}=2M/\mu$; /* upper bound on the number of UEs $N$ */\\
$N=N^{\rm{lower}}+\lceil (N^{\rm{upper}}-N^{\rm{lower}})/2 \rceil$; \\
\textbf{while} $N^{\max}=0$ \\
\qquad \textbf{if} $\varepsilon(N)$ has not been calculated\\
\qquad\qquad calculate $\varepsilon(N)$ according to \eqref{eq-reliability-delay}\\
\qquad\qquad \textbf{if} $\varepsilon(N) = \varepsilon ^{\max } $ \\
\qquad\qquad\qquad $N^{\max}=N$; \\
\qquad\qquad \textbf{else if} $\varepsilon(N) > \varepsilon ^{\max }$ \\
\qquad\qquad\qquad $N^{\rm{upper}} = N$; /* reduce the number of UEs */ \\
\qquad\qquad \textbf{else} \\
\qquad\qquad\qquad $N^{\rm{lower}} = N$; /* increase the number of UEs */ \\
\qquad\qquad \textbf{end if} \\
\qquad\qquad $N=N^{\rm{lower}}+\lceil (N^{\rm{upper}}-N^{\rm{lower}}/2 \rceil$; \\
\qquad \textbf{else}\\
\qquad\qquad $N^{\max}=N$;\\
\qquad \textbf{end if}\\
\textbf{end while} \\
\hline
\end{tabular*}
\end{table}

\par
Next, we find the optimal number of receiving antennas $M^*$ for maximizing the system energy efficiency and satisfying the constraint $\varepsilon \leq \varepsilon ^{\max }$ simultaneously, when the other parameters are fixed.
In this case, the optimization problem is with respect to the number of receiving antennas $M$.
Therefore, we rewrite the delay-constrained reliability $\varepsilon$ as $\varepsilon(M)$, and rewrite the system energy efficiency $\eta$ as $\eta(M)$.
The system energy efficiency optimization problem under the constraint of the reliability and delay QoS is formulated as
\begin{equation}
\label{eq-OP-M}
\renewcommand{\arraystretch}{1.1}
\begin{array}{l}
\mathop {\max }\limits_{M} \;\eta(M)\\
\text{s.t.}\quad \varepsilon(M) \leq \varepsilon ^{\max }, M \in \mathbb{Z}^{+}.\\
\end{array}
\end{equation}
The basic idea of solving \eqref{eq-OP-M} is similar to the one for \eqref{eq-OP-N}. Obviously, $\varepsilon(M)$ is a decreasing function with respect to $M$. Thus, we first utilize the dichotomy algorithm to search for the minimum number of receiving antennas that can satisfy the reliability and delay QoS requirement $\varepsilon(M) \leq \varepsilon ^{\max}$. This minimum number of receiving antennas is defined as $M^{\min}=\min\{M|\varepsilon(M) \leq \varepsilon ^{\max }, M \in \mathbb{Z}^{+}\}$. Define $M^{\max}$ as the maximum allowable number of receiving antennas equipped at the BS. $M^{\max}$ is limited by the receiver hardware of BS. Then, the exhaustive search method is utilized to find $M^*$ in the interval that is from $M^{\min}$ to $M^{\max}$. The total computational complexity of the search processes including the dichotomy algorithm and the exhaustive search method is $O(\log_2 {(M^{\max}-1)})+O(M^{\max}-M^{\min})$.

\section{Simulation results and discussions}
\label{sec-simulation}
\begin{figure}[!t]
	\centering
	\includegraphics[width=2.5 in]{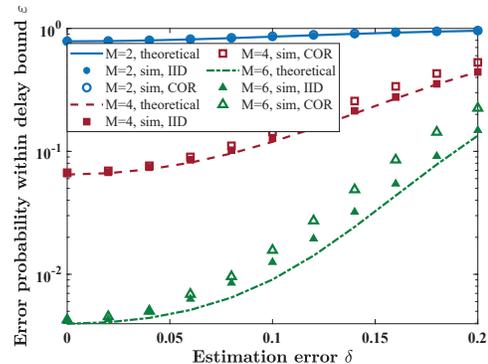}
	\caption{The theoretical analysis and numerical simulation results of the error probability within the delay bound versus the channel estimation error. Assume $\xi = -92.4$ dBm, $K_{\text{rep}}=1$, $K_{\text{F}}=1$, $\mu = 0.6$ and $N = 10$.}
	\label{fig-unreliability-delta}
\end{figure}

\begin{figure}[!t]
	\centering
	\includegraphics[width=2.6 in]{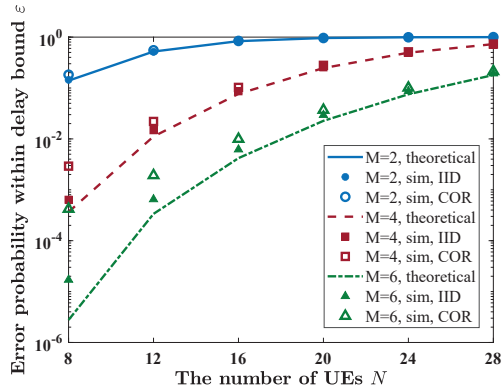}
	\caption{The theoretical analysis and numerical simulation results of the error probability within the delay bound versus the number of UEs. Assume $\xi = -92.4$ dBm, $K_{\text{rep}}=1$, $K_{\text{F}}=1$, $\mu = 0.4$ and $\delta=0.1$.}
	\label{fig-unreliability-N}
\end{figure}

\begin{figure}[!t]
	\centering
	\includegraphics[width=2.6 in]{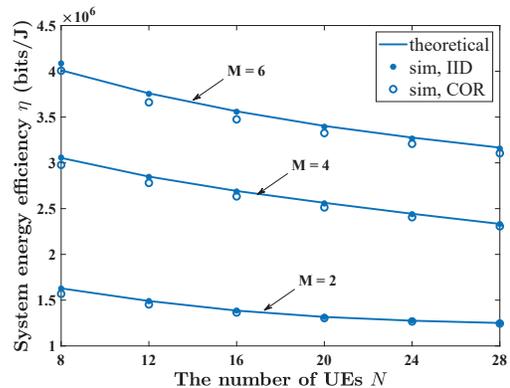}
	\caption{The theoretical results and the numerical simulation results of the system energy efficiency versus the number of UEs. Assume $\xi = -92.4$ dBm, $K_{\text{rep}}=1$, $K_{\text{F}}=1$, $\mu = 0.6$ and $\delta=0.1$.}
	\label{fig-EE-N}
\end{figure}

\begin{figure}[!t]
	\centering
	\includegraphics[width=2.6 in]{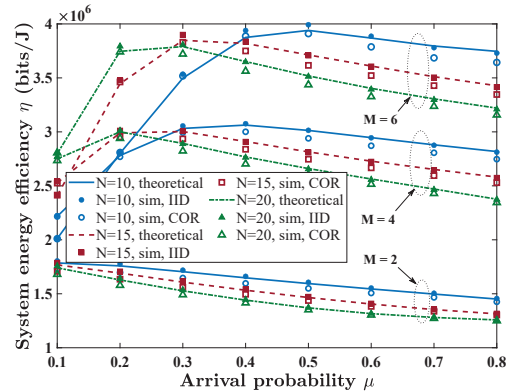}
	\caption{The theoretical results and numerical simulation results of the system energy efficiency versus the arrival probability. Assume $\xi = -92.4$ dBm, $K_{\text{rep}}=1$, $K_{\text{F}}=1$, and $\delta=0.1$.}
	\label{fig-EE-mu}
\end{figure}

\begin{figure*}[!t]
\centering
\subfloat[]{\includegraphics[width=2.6 in]{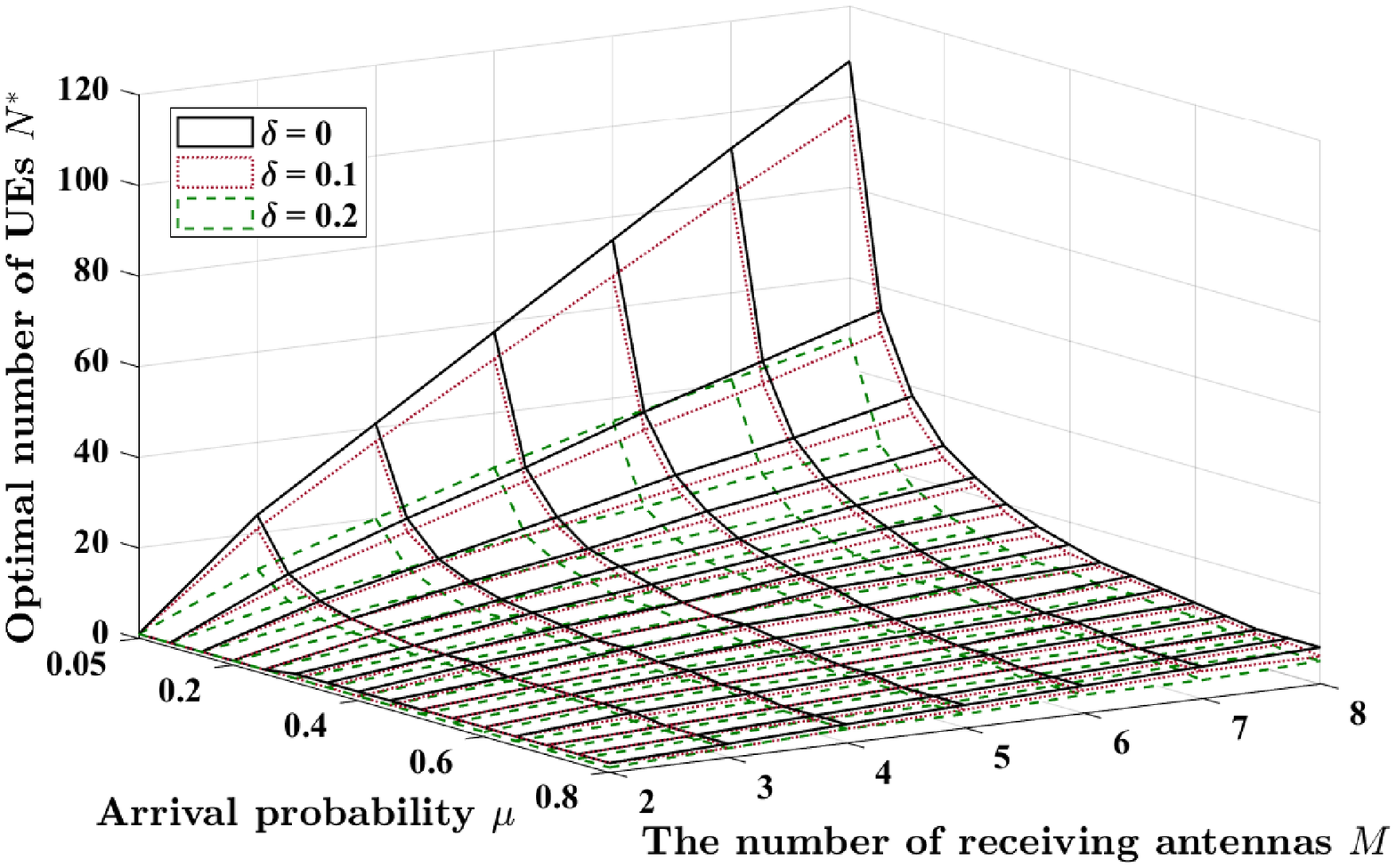}
\label{fig-optimalN-mu-N}}\hspace{10mm}
\subfloat[]{\includegraphics[width=2.6 in]{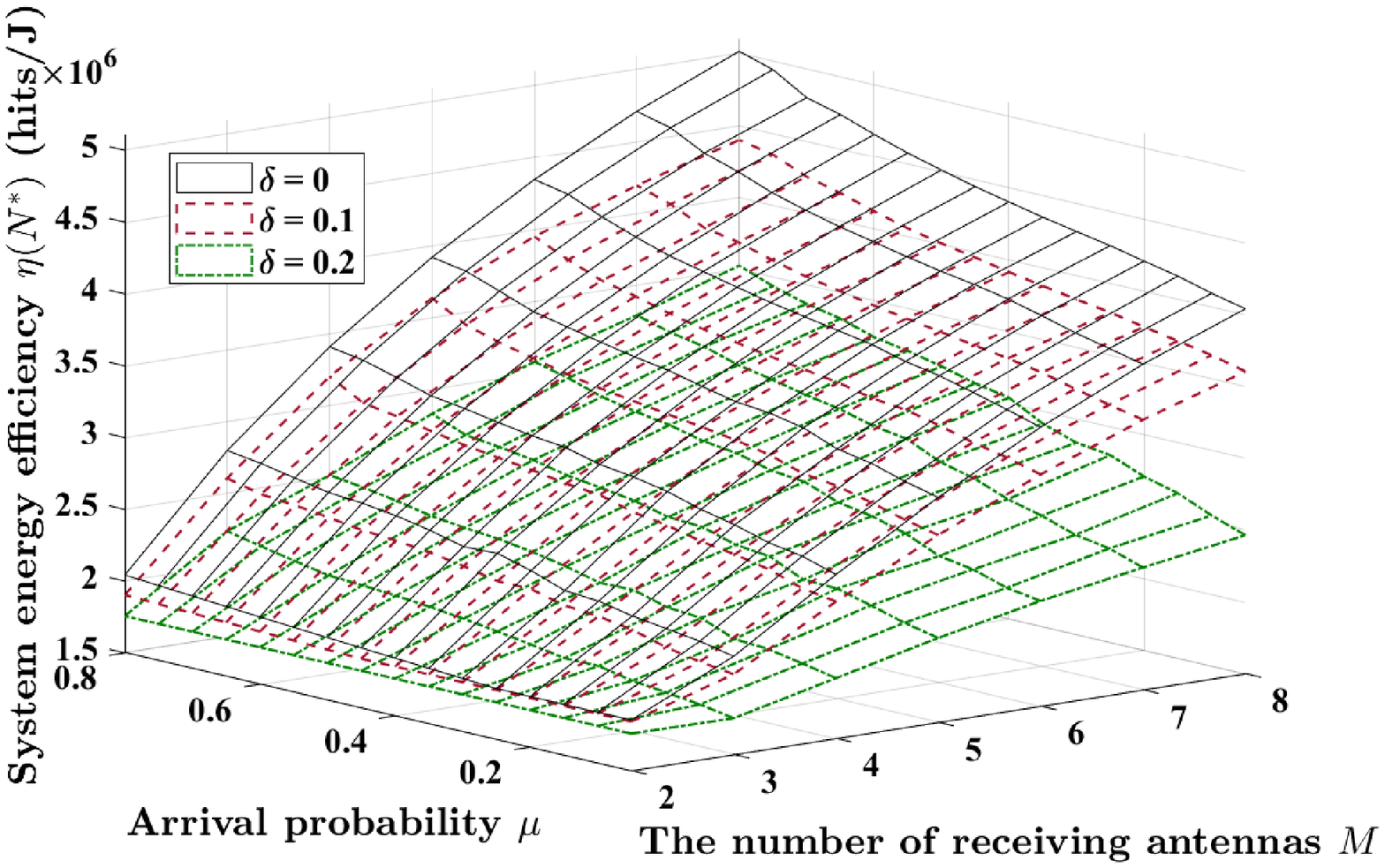}%
\label{fig-optimalN-mu-EE}}
\caption{The optimal number of UEs and the optimal system energy efficiency versus the arrival probability, while assuming $\xi = -90$ dBm, $K_{\text{rep}}=1$ and $K_{\text{F}}=1$. (a) The optimal number of UEs (i.e., $N^*$). (b) The system energy efficiency (i.e., $\eta(N^*)$).}
\label{fig-optimalN-mu}
\end{figure*}

\begin{figure*}[!t]
\centering
\subfloat[]{\includegraphics[width=2.6 in]{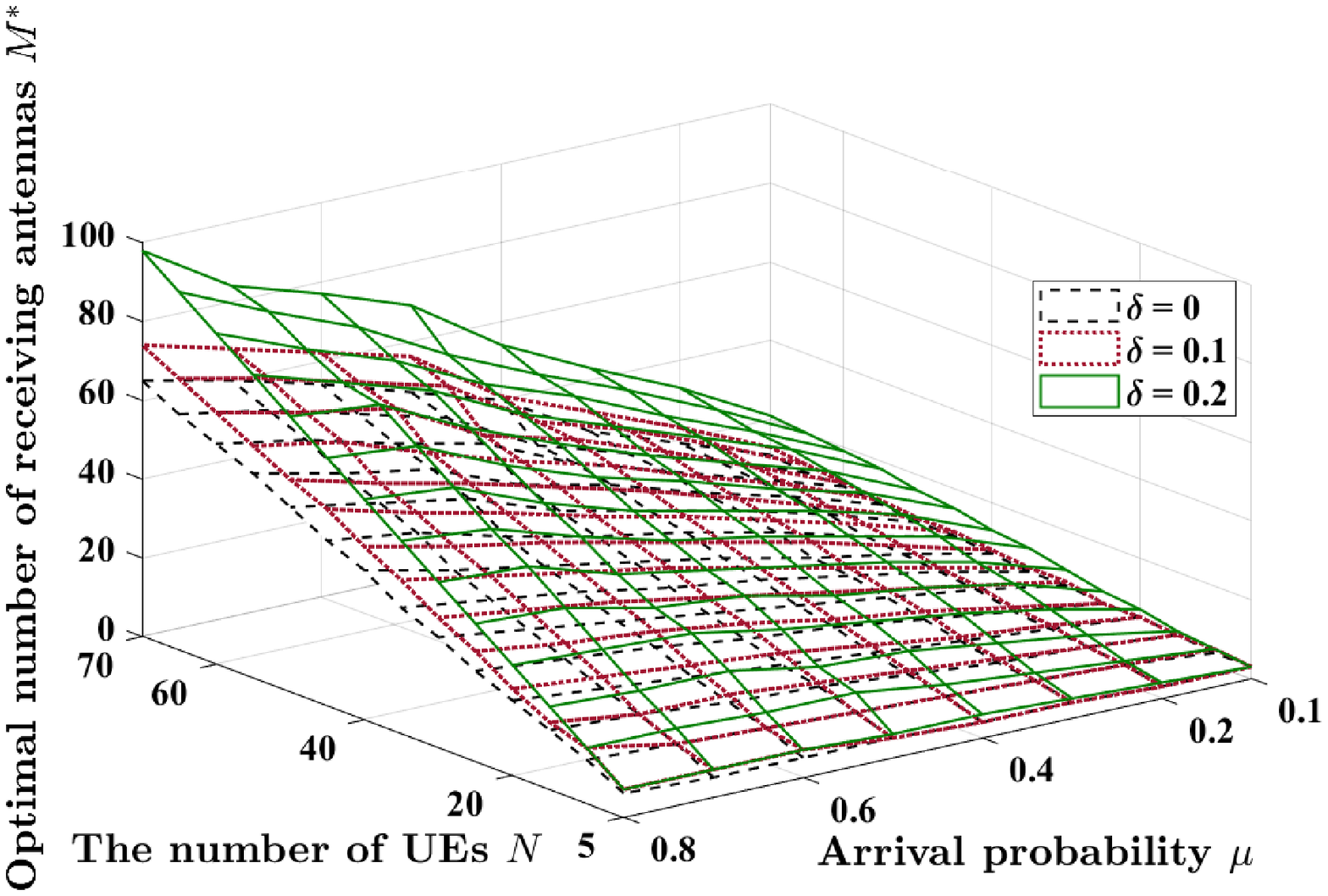}%
\label{fig-optimalM-N-M}}\hspace{15mm}
\subfloat[]{\includegraphics[width=2.6 in]{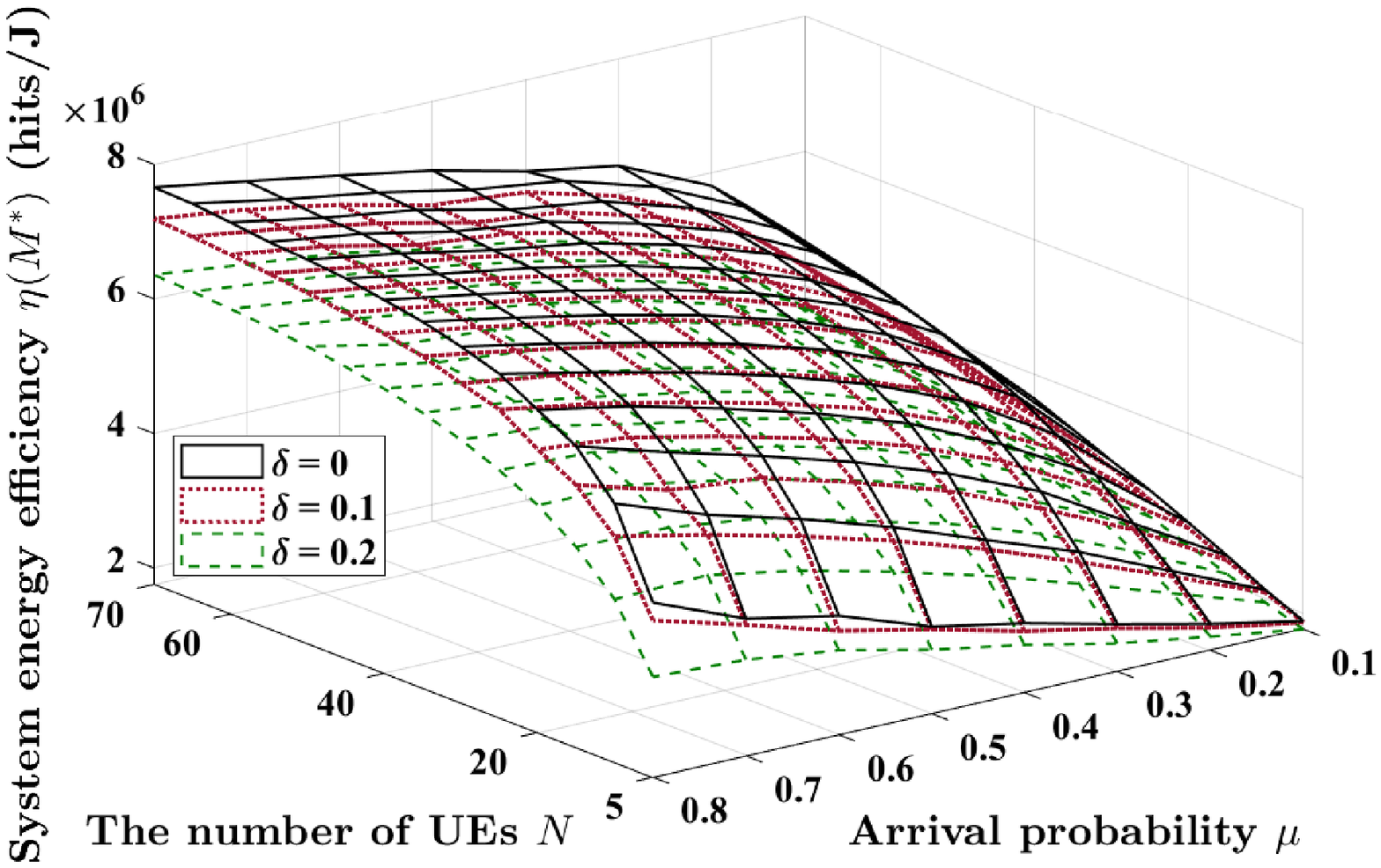}%
\label{fig-optimalM-N-EE}}
\caption{The optimal number of receiving antennas and the optimal system energy efficiency versus the arrival probability, while assuming $\xi = -90$ dBm, $K_{\text{rep}}=1$, $K_{\text{F}}=1$, and $M^{\max}=2N$. (a) The optimal number of receiving antennas (i.e., $M^*$). (b) The system energy efficiency (i.e., $\eta(M^*)$).}
\label{fig-optimalM-N}
\end{figure*}

In our simulations, the BS is equipped with $M$ receiving antennas, the circuit power consumed per receiving antenna is $P^{\rm{ca}}=17$ dBm, and the transmit power of the BS is $P^{\text{BS-tx}}=30$ dBm.
$N$ UEs are uniformly located with the distance from the BS as $50$ m $\sim 150$ m away from the BS, and each UE is equipped with one transmitting antenna.
The noise power spectral density  $N_0$ is assumed to be $-174$ dBm/Hz.
In addition, the subcarrier spacing (SCS) is $30$ kHz, a single OFDM symbol occupies $1/28$ ms, and each S-TTI is equal to $2$ OFDM symbols, i.e., $\tau=1/14$ ms. We assume that $30$ subcarriers are dedicated to the uplink URLLC traffic, i.e., $B=900$ kHz.
The packet size of the URLLC traffic is set to $\beta = 160$ bits \cite{3GPP-38802}.
The delay bound $D^{\max}$ is $1$ ms in the scenarios of Fig. \ref{fig-unreliability-delta} to Fig. \ref{fig-optimalM-N}.
Additionally, we assume that the reliability requirement is $99.999\%$, i.e., $\varepsilon^{\max} = 10^{-5}$.

\begin{figure*}[!t]
	\centering
	\subfloat[]{\includegraphics[width=2.6 in]{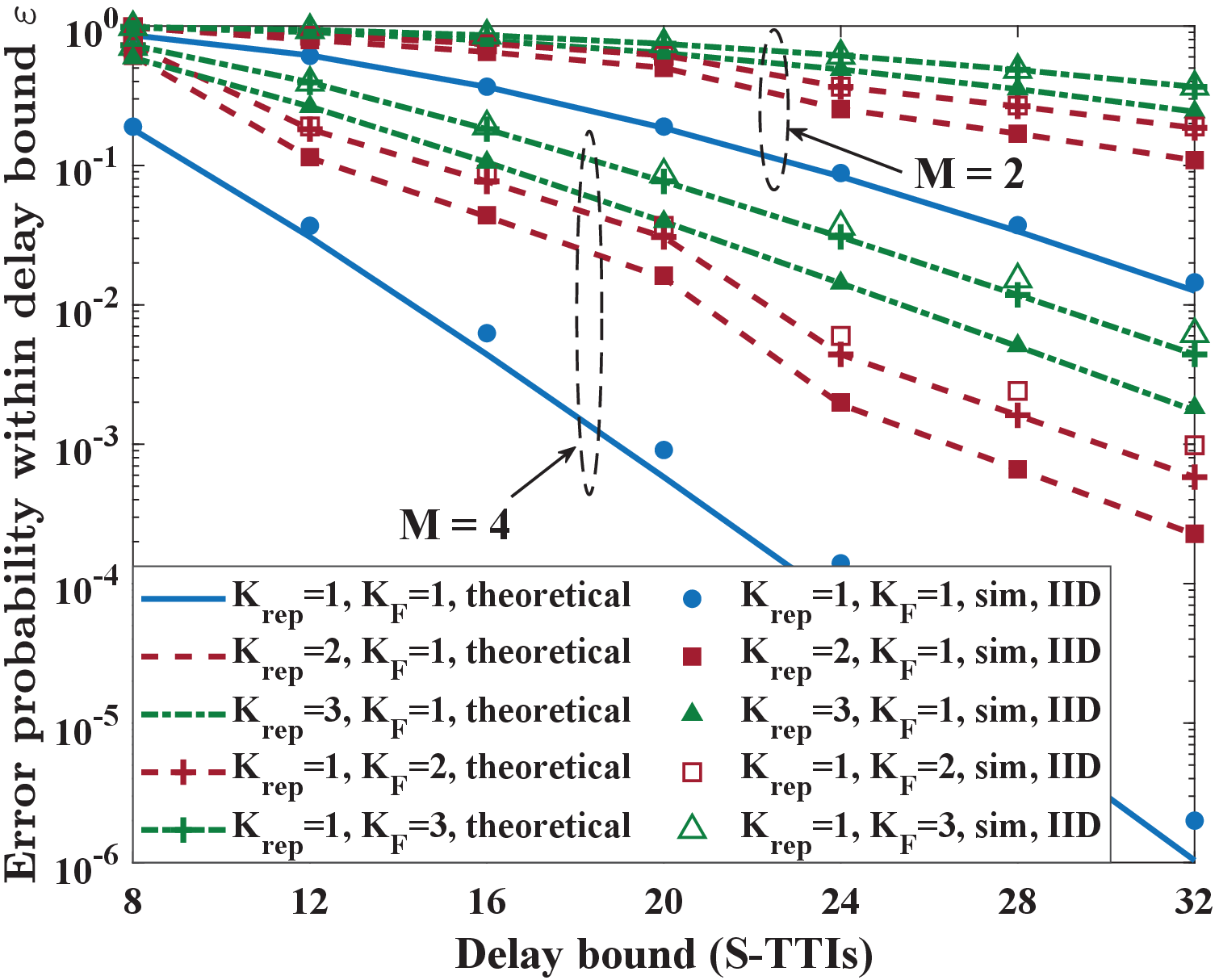}%
		\label{fig-unreliability-Krep-a}}\hspace{15mm}
	\subfloat[]{\includegraphics[width=2.6 in]{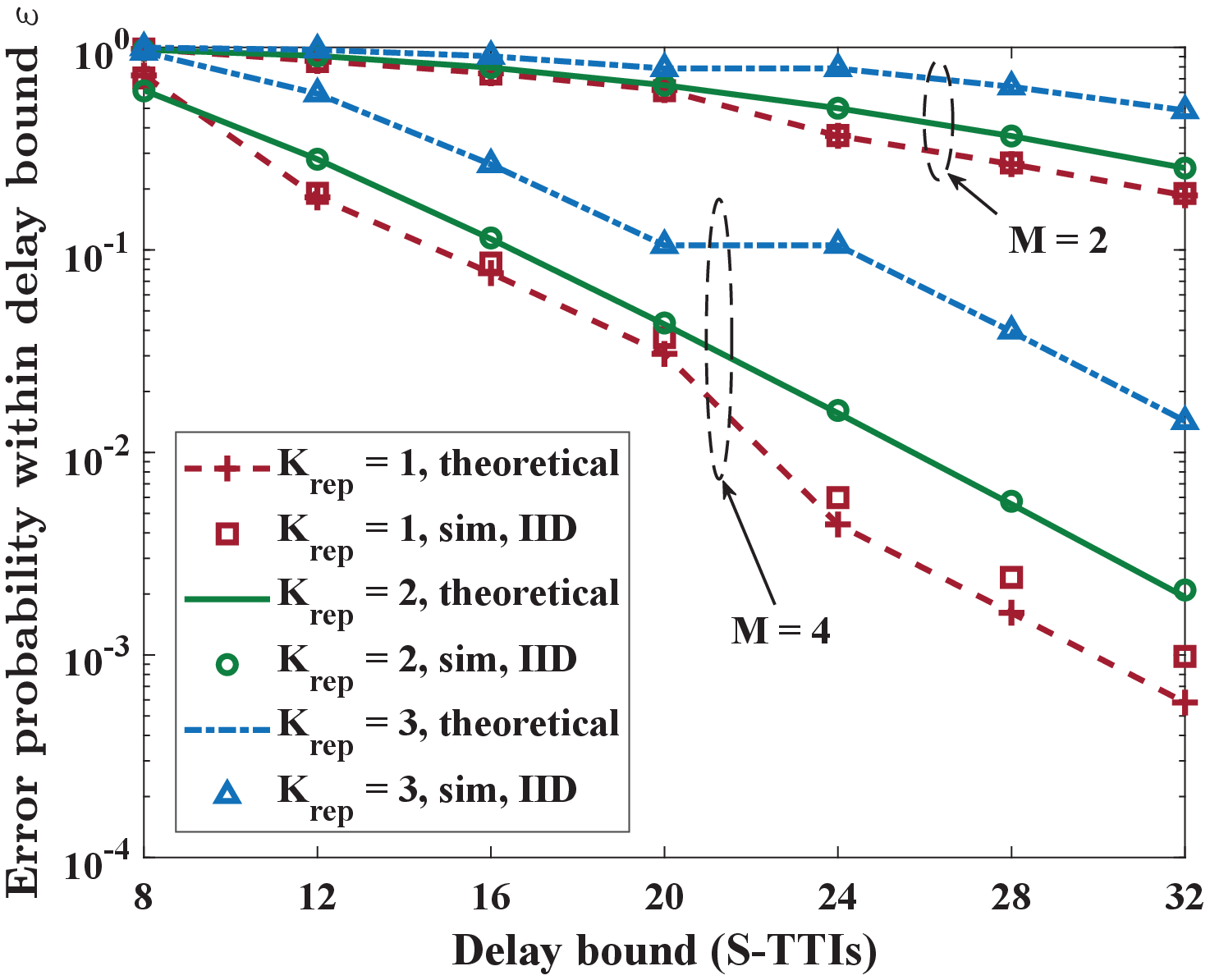}%
		\label{fig-unreliability-Krep-b}}
	\caption{The delay-bound-constrained error probability $\varepsilon$ versus the delay bound $L$ under different numbers of the receiving antennas $M$. Assume $\xi = -92.4$ dBm, $\mu = 0.5$ and $N = 14$. (a) The results under different numbers of the repetitions $K_{\text{rep}}$ and the given waiting time of the feedback $K_{\text{F}} = 1$, as well as the results under different values of $K_{\text{F}}$ and the given $K_{\text{rep}} =1$  . (b) The results under different numbers of the repetitions $K_{\text{rep}}$ and the given $K_{\text{F}}=2$.}
	\label{fig-unreliability-Krep}
\end{figure*}

\par
In this section, we first validate the absorbing-Markov-chain-based analysis by comparing the theoretical results with the numerical simulation results, as shown in Fig. \ref{fig-unreliability-Krep} $\sim$ Fig.~\ref{fig-EE-mu}. Then we demonstrate the energy efficiency of the system in the case of the optimal number of UEs and the optimal number of receiving antennas, as shown in Fig.~\ref{fig-optimalN-mu} $\sim$ Fig.~\ref{fig-optimalM-N}.
In this paper, all numerical results are based on the $1 \times 10^6$ times Monte Carlo trials performed using Matlab.
We simulate the Bernoulli arrivals of URLLC traffic, the access-probability-adaptive grant-free transmissions with the $K$-repetition scheme of the UEs, as well as the ZF-MUD algorithm and the MRC algorithm of the BS.
In Monte Carlo simulations, the PHY transmission rate is characterized by the maximal achievable rate formula in the finite block-length regime, as depicted in \eqref{eq-rate-bits}. In addition, the block fading is considered.
In our results, ``sim, IID'' represents the numerical simulation results of the independent scenario where the channel coherence time is assumed to be one S-TTI, and the channel gain varies independently among S-TTIs. This assumption is also followed by our theoretical analysis. In practice, the channel coherence time is usually larger than one S-TTI \cite{URLLC-Yang-crosslayer}. Thus, for testing the assumption of the theoretical analysis, we also present the numerical simulation results tagged as ``sim, COR'' to represent the results in the correlated scenario. In this scenario, the channel coherence time is assumed to be one slot, so the channel gain remains unchanged in the multiple transmissions of one data packet.

\par
Fig. \ref{fig-unreliability-Krep} shows the delay-bound-constrained error probability $\varepsilon$ versus the delay bound $L$ in units of S-TTI under different numbers of the receiving antennas $M$. More specifically, Fig. \ref{fig-unreliability-Krep}\subref{fig-unreliability-Krep-a} presents the results under different numbers of the repetitions $K_{\text{rep}}$ and the given waiting time of the feedback $K_{\text{F}}=1$, as well as the results under different values of $K_{\text{F}}$ and the given $K_{\text{rep}} =1$. Fig. \ref{fig-unreliability-Krep}\subref{fig-unreliability-Krep-b} shows the results under different numbers of the repetitions $K_{\text{rep}}$ and the given $K_{\text{F}}=2$. From Fig. \ref{fig-unreliability-Krep}, it can be seen that the theoretical analysis results match the numerical results very well, except in the cases of $M=4$, $K_{\text{rep}}=1$, $K_{\text{F}}=1, 2$ and $L \geq 24$ (i.e., $D^{\max} \geq 1.71$ ms). The reason is explained as follows.
Recall that $\varepsilon$ is derived based on the absorbing Markov chain formulated.
When analysing the transition probabilities of the Markov chain, we neglect an unlikely outage event, where the same group of residual contention UEs with strong spatially correlated channel fading may retransmit simultaneously. Although this case is fairly rare, it does have some negative impact on URLLCs. 
As shown in Fig. \ref{fig-unreliability-Krep}\subref{fig-unreliability-Krep-a}, in the case of $K_{\text{F}}=1$, the error probability decreases with the delay bound, but increases when $K_{\text{rep}}$ grows. Similar trend can also be found in Fig. \ref{fig-unreliability-Krep}\subref{fig-unreliability-Krep-b} for $K_{\text{F}}=2$. The reason is simple: The post-processing SNR of one transmission improves with increasing $K_{\text{rep}}$ at the cost of reducing the number of retransmissions. Additionally, the ZF-MUD algorithm is superior to the MRC algorithm in terms of improving SNR.
This indicates that retransmissions are more beneficial for improving the reliability of the access-probability-adaptive grant-free access. From Fig.~\ref{fig-unreliability-Krep}\subref{fig-unreliability-Krep-a}, we also observe that the error probability in the case of ``$K_{\text{rep}}=2$ and $K_{\text{F}}=1$'' is smaller than the one in the case of ``$K_{\text{rep}}=1$ and $K_{\text{F}}=2$''. Similar results can be found in the case of ``$K_{\text{rep}}=3$ and $K_{\text{F}}=1$'' and the case of ``$K_{\text{rep}}=1$ and $K_{\text{F}}=3$''. This indicates that the $K$-repetition scheme is more beneficial for improving the reliability when the round-trip-delay equal to $K_{\text{rep}}+K_{\text{F}}$ [S-TTIs] is fixed. 

\par
Fig. \ref{fig-unreliability-delta} shows the delay-bound-constrained error probability $\varepsilon$ versus the channel error $\delta$ under different numbers of receiving antennas $M$.
Fig. \ref{fig-unreliability-N} shows the delay-bound-constrained error probability $\varepsilon$ versus the number of UEs $N$ under different $M$.
In Fig. \ref{fig-unreliability-delta} and Fig. \ref{fig-unreliability-N}, the gap between our theoretical results and the numerical results in correlated scenarios (i.e., the results tagged as ``sim, COR'') becomes larger with the increase of $M$. This is for the following reasons. When $M$ becomes larger, the group of residual contention UEs, which suffer not only strong spatially correlated but also temporally correlated channel fading, are more likely to transmit simultaneously.
We will investigate this phenomenon more deeply in the future for supporting URLLCs more efficiently. 
From Fig. \ref{fig-unreliability-delta}, it can also be seen that the channel estimation error deteriorates the reliability more when $M$ becomes larger. It indicates that the channel estimation is more important for large MIMO systems.
Fig. \ref{fig-unreliability-N} also demonstrates that more receiving antennas are needed for satisfying the delay and reliability QoS requirements of URLLC traffic in the scenario where $N > 12$ and $\mu = 0.4$.

\par
Fig. \ref{fig-EE-N} shows the system energy efficiency $\eta$ versus the number of UEs (i.e., $N$) under different numbers of the receiving antennas (i.e., $M$).
Fig. \ref{fig-EE-mu} shows the system energy efficiency $\eta$ versus the arrival probability $\mu$ under different numbers of UEs. They both demonstrate that the theoretical results match well with the numerical simulation results, even in the correlated scenarios.
As shown in Fig. \ref{fig-EE-N}, the system energy efficiency degrades with the number of UEs for the given number of receiving antennas, when considering $\mu=0.6$.
From Fig. \ref{fig-EE-mu}, it can be seen that the system energy efficiency first improves and then degrades with the arrival probability when $M=4$ and $M=6$. However, the system energy efficiency decreases with the arrival probability when $M=2$. It is also found that the system energy efficiency decreases with the arrival probability after $\mu \geq M/N$ whatever $M$ is. It indicates that even more packets may be transmitted successfully with the increase of $\mu$, but much more energy is cost after $\mu \geq M/N$, which leads to deterioration of the system energy efficiency.

\par
Fig. \ref{fig-optimalN-mu}\subref{fig-optimalN-mu-N} and Fig. \ref{fig-optimalN-mu}\subref{fig-optimalN-mu-EE} show the optimal number of UEs (i.e., $N^*$) and the optimal system energy efficiency (i.e., $\eta(N^*)$) versus the arrival probability $\mu$ and the number of receiving antennas $M$, respectively.
It can be seen from Fig. \ref{fig-optimalN-mu}\subref{fig-optimalN-mu-N} that the optimal number of UEs decreases with the arrival probability. The rate of descent is sharp first, and then becomes moderate after $\mu>0.2$.
Additionally, the optimal number of UEs has a nearly linear increase with the number of receiving antennas, and the growth rate is faster when the arrival probability is smaller. 
From Fig. \ref{fig-optimalN-mu}\subref{fig-optimalN-mu-EE}, it can be seen that the optimal system energy efficiency increases with the arrival probability, and it increases faster in the region of large $M$. It can also be seen that the optimal system energy efficiency increases with the number of receiving antennas. In addition, as shown in Fig.~\ref{fig-optimalN-mu}, the results about $N^*$ and $\eta(N^*)$ of $\delta=0.1$ are close to that of $\delta=0$, while the results of $\delta=0.2$ are far lower than that of $\delta=0$. It indicates that the channel estimation errors of $\delta \leq 0.1$ can be tolerated.

\par
Fig. \ref{fig-optimalM-N}\subref{fig-optimalM-N-M} and Fig. \ref{fig-optimalM-N}\subref{fig-optimalM-N-EE} show the optimal number of receiving antennas (i.e., $M^*$) and the optimal system energy efficiency (i.e., $\eta(M^*)$) versus the number of UEs $N$ and the arrival probability $\mu$, respectively.
As shown in Fig. \ref{fig-optimalM-N}\subref{fig-optimalM-N-M}, more receiving antennas are needed when either the number of UEs or the arrival probability increases for supporting energy-efficient uplink URLLC.
Additionally, the optimal number of receiving antennas increases almost linearly with the number of UEs, and the increasing rate is larger in the case of larger arrival probability.
It demonstrates that the massive MIMO is a feasible way to support energy-efficient URLLC in the large-scale system with frequent arrivals.
From Fig. \ref{fig-optimalM-N}\subref{fig-optimalM-N-EE}, it can be seen that the optimal system energy efficiency increases as the arrival probability or the number of UEs increases. Furthermore, it is observed that the optimal system energy efficiency seems to converge when the number of UEs and the arrival probability both become large.
This indicates that the access-probability-adaptive grant-free transmission with ZF-MUD is robust to the large-scale system if the number of receiving antennas can be optimized according to the system energy efficiency.

\section{Conclusion}
\label{sec-conclusion}
In this paper, we dedicated to handling the issue that how to achieve uplink URLLC in the ZF-MUD-aided grant-free access with high energy efficiency.
We proposed the absorbing-Markov-chain-based analysis framework to derive the delay-constrained reliability and system energy efficiency for the ZF-MUD-aided grant-free access system under the assumption of Rayleigh fading with channel estimation error.
In contrast to prior works, the ZF-MUD technique, $K$-repetition and ACK-enabled retransmission within delay bound were all considered.
The simulation results showed that one repetition is enough for the ZF-MUD-aided access-probability-adaptive grant-free transmissions.
Furthermore, we utilized the dichotomy algorithm and the exhaustive search method to find the optimal number of UEs and the optimal number of receiving antennas that maximize the system energy efficiency, while satisfying the reliability and delay QoS at the same time.
Our simulation results also showed that the massive MIMO is a feasible way to support uplink large-scale URLLC systems.
\par
The framework of our theoretical analysis relies on the assumptions on the channel model and the arrival processes, which are not exactly the same as those in practical systems. The deep reinforcement learning may be a feasible way to handle this model mismatch problem~\cite{6G-learning}. Thus, we will study how to integrate the framework of theoretical analysis into the model-free deep reinforcement learning methods for URLLCs in the future.


%

\appendices
\section{Proof of Theorem \ref{lemma-outage}}
\label{app-outage-probability}
In this paper, the transmission of a packet for a UE is regarded as successful if its transmission rate is not smaller that the threshold $\beta/\tau$. To derive $\rho(N')$ (i.e., $\Pr\left\{ \displaystyle R(\gamma(n'|N')) < \beta/\tau \right\}$), considering the feature of short packet transmission, we refer to \eqref{eq-rate-bits}. In the regime of $\gamma(n'|N') \geq 10$, we have $V(\gamma(n'|N')) \approx 1$. Then
\begin{equation}
\label{eq-rate-bits-app}
R(\gamma(n'|N')) \approx
B {\displaystyle \left[ {{{\log }_2}\left( {1 + {\gamma(n'|N')}} \right) - \sqrt {\frac{{{\log }_2}e}{\tau B}}f_{\rm{Q}}^{ - 1}(\varepsilon^{\rm{B}}) } \right] }.
\end{equation}
Hence, the conditional outage probability of a UE is reformulated as
\begin{equation}
\label{eq-outage-pr}
\begin{array}{l}
\rho(N') = \displaystyle \Pr\left\{ R(\gamma(n'|N')) <\frac{\beta}{\tau} \right\}\\
\mathop  = \limits^{(a)} \displaystyle \Pr\left\{ B {\displaystyle \left[ {{{\log }_2}\left( {1 + {\gamma(n'|N')}} \right) - \sqrt {\frac{{{\log }_2}e}{\tau B}}f_{\rm{Q}}^{ - 1}(\varepsilon^{\rm{B}}) } \right] }<\frac{\beta}{\tau} \right\}\\
= \displaystyle \Pr\left\{ \gamma(n'|N') < 2^{\displaystyle \sqrt {\frac{{{\log }_2}e}{\tau B}}f_{\rm{Q}}^{ - 1}(\varepsilon^{\rm{B}})+\frac{\beta}{\tau B}}-1 \right\}.
\end{array}
\end{equation}
In \eqref{eq-outage-pr}, the equation (a) is obtained by substituting \eqref{eq-rate-bits-app} to $R(\gamma(n'|N'))$. For clarity of presentation, define
\[\Omega=\displaystyle \sqrt {\frac{{{\log }_2}e}{\tau B}}f_{\rm{Q}}^{ - 1}(\varepsilon^{\rm{B}})+\frac{\beta}{\tau B} ,\]
then \eqref{eq-outage-pr} is rewritten as
\begin{equation}
\label{eq-putage-pr-concise}
\rho(N') = \displaystyle \Pr\left\{ \gamma(n'|N') < 2^{\Omega}-1 \right\}.
\end{equation}
From \eqref{eq-zf-sinr}, we have
\begin{equation}
\label{eq-outage-pr-sinr}
\begin{aligned}
& \displaystyle \Pr\left\{ \gamma(n'|N') < 2^{\Omega}-1 \right\} \\
& = \displaystyle \Pr\left\{ \sum\limits_{l=1}^{K_{\text{rep}}} \frac{\xi}{(N_0 B+\xi \delta^2 N') \left[ (\hat{\bm{H}}^{\rm{H}}\hat{\bm{H}})^{-1} \right]_{n',n'}} < 2^{\Omega}-1 \right\}\\
& = \displaystyle \Pr\left\{ \sum\limits_{l=1}^{K_{\text{rep}}} \frac{1}{\left[ (\hat{\bm{H}}^{\rm{H}}\hat{\bm{H}})^{-1} \right]_{n',n'}} < \frac{(N_0 B + \xi \delta^2 N')(2^{\Omega}-1)}{\xi} \right\}.
\end{aligned}
\end{equation}
Please note that $\hat{\bm{H}}$ is assumed to be IID across multiple S-TTIs in our analysis. From \eqref{eq-channel-error}, it can be found that $\hat{\bm{H}}$ follows ${\cal{CN}}(0, (\delta^2+1){\bm{I}}_{M})$. Then, $ \hat{\bm{H}}^{\rm{H}}\hat{\bm{H}} $ follows the Wishart distribution $W_{N'}(M, (\delta^2+1){\bm{I}}_{N'})$. According to Theorem 3.2.11 in \cite{book-MST}, ${\raise0.7ex\hbox{$1$} \!\mathord{\left/
 {\vphantom {1 \left[ {\left( \hat{\bm{H}}^{\rm{H}}\hat{\bm{H}} \right)}^{-1} \right]_{n', n'}}}\right.\kern-\nulldelimiterspace}
\!\lower0.7ex\hbox{$\left[ {\left( \hat{\bm{H}}^{\rm{H}}\hat{\bm{H}} \right)}^{-1} \right]_{n', n'}$}}$ for $1 ~\leq~n'~\leq~N'$, follows the Gamma distribution with the shape $M-N'+1$ and scale $\delta^2+1$. Hence, the sum of the IID Gamma distributed random variables (i.e., $\displaystyle\sum\limits_{l=1}^{K_{\text{rep}}} \frac{1}{\left[ (\hat{\bm{H}}^{\rm{H}}\hat{\bm{H}})^{-1} \right]_{n',n'}}$) follows the Gamma distribution with the shape $K_{\text{rep}}(M-N'+1)$ and scale $\delta^2+1$.
Accordingly, the conditional outage probability of a UE is
\begin{equation}
\label{eq-outage-pr-sinr-gamma}
\begin{aligned}
	& \displaystyle \Pr\left\{ \gamma(n'|N') < 2^{\Omega}-1 \right\} \\
	& = \frac{\displaystyle f_{\gamma}\left( K_{\text{rep}}(M-N'+1), \frac{(N_0 B + \xi \delta^2 N')(2^{\Omega}-1)}{\xi(\delta^2+1)} \right)}{f_{\Gamma}(K_{\text{rep}}(M-N'+1))},
\end{aligned}
\end{equation}
where $f_{\gamma}(\cdot,\cdot)$ is the lower incomplete Gamma function, and $f_{\Gamma}(\cdot)$ is the Gamma function.



\ifCLASSOPTIONcaptionsoff
  \newpage
\fi



%
%
%

\bibliographystyle{IEEEtran}
\bibliography{IEEEabrv,urllc-mimo-references}

\begin{thebibliography}{10}
\providecommand{\url}[1]{#1}
\csname url@samestyle\endcsname
\providecommand{\newblock}{\relax}
\providecommand{\bibinfo}[2]{#2}
\providecommand{\BIBentrySTDinterwordspacing}{\spaceskip=0pt\relax}
\providecommand{\BIBentryALTinterwordstretchfactor}{4}
\providecommand{\BIBentryALTinterwordspacing}{\spaceskip=\fontdimen2\font plus
\BIBentryALTinterwordstretchfactor\fontdimen3\font minus
  \fontdimen4\font\relax}
\providecommand{\BIBforeignlanguage}[2]{{%
\expandafter\ifx\csname l@#1\endcsname\relax
\typeout{** WARNING: IEEEtran.bst: No hyphenation pattern has been}%
\typeout{** loaded for the language `#1'. Using the pattern for}%
\typeout{** the default language instead.}%
\else
\language=\csname l@#1\endcsname
\fi
#2}}
\providecommand{\BIBdecl}{\relax}
\BIBdecl

\bibitem{URLLC-Bennis}
M.~{Bennis}, M.~{Debbah}, and H.~V. {Poor}, ``Ultrareliable and low-latency
  wireless communication: Tail, risk, and scale,'' \emph{Proc. {IEEE}}, vol.
  106, no.~10, pp. 1834--1853, Oct. 2018.

\bibitem{URLLC-Power}
H.~{Ren}, C.~{Pan}, Y.~{Deng}, M.~{Elkashlan}, and A.~{Nallanathan}, ``Joint
  power and blocklength optimization for {URLLC} in a factory automation
  scenario,'' \emph{{IEEE} Trans. Wireless Commun.}, vol.~19, no.~3, pp.
  1786--1801, Dec. 2020.

\bibitem{3GPP-38913}
3GPP, ``Study on scenarios and requirements for next generation access
  technologies,'' Tech. Spec. Group Radio Access Network, tech. rep. 38.913
  v15.0.0, Release 15, Jun. 2018.

\bibitem{URLLC-LTE-VTC-2017}
C.~Wang, Y.~Chen, Y.~Wu, and L.~Zhang, ``Performance evaluation of grant-free
  transmission for uplink {URLLC} services,'' in \emph{{IEEE} Veh. Technol.
  Conf. (VTC Spring)}, Sydney, NSW, Australia, Jun. 2017, pp. 1--6.

\bibitem{3GPP-38802}
3GPP, ``Study on new radio access technology physical layer aspects,'' Tech.
  Spec. Group Radio Access Network, tech. rep. 38.802 v14.2.0, Release 14, Sep.
  2017.

\bibitem{grant-free-VTC}
A.~Azari, M.~I. Hossain, and J.~I. Markendahl, ``{RACH} dimensioning for
  reliable {MTC} over cellular networks,'' in \emph{{IEEE} Veh. Technol. Conf.
  (VTC Spring)}, Sydney, NSW, Australia, Jun. 2017, pp. 1--5.

\bibitem{grant-free-ACCESS}
G.~Berardinelli, N.~H. Mahmood, R.~Abreu, T.~Jacobsen, K.~Pedersen, I.~Z.
  Kovács, and P.~Mogensen, ``Reliability analysis of uplink grant-free
  transmission over shared resources,'' \emph{IEEE Access}, vol.~6, pp.
  23\,602--23\,611, Apr. 2018.

\bibitem{Shaoshi2015}
S.~Yang and L.~Hanzo, ``Fifty years of {MIMO} detection: The road to
  large-scale {MIMO}s,'' \emph{{IEEE} Commun. Surveys Tuts.}, vol.~17, no.~4,
  pp. 1941--1988, 4th Quart. 2015.

\bibitem{Linlin-Aloha-WC}
L.~Zhao, X.~Chi, and S.~Yang, ``Optimal {ALOHA}-like random access with
  heterogeneous {QoS} guarantees for multi-packet reception aided visible light
  communications,'' \emph{{IEEE} Trans. Wireless Commun.}, vol.~15, no.~11, pp.
  7872--7884, Nov. 2016.

\bibitem{URLLC-2017-GLOBECOM-grant-free}
A.~Azari, P.~Popovski, G.~Miao, and C.~Stefanovic, ``Grant-free radio access
  for short-packet communications over {5G} networks,'' in \emph{Proc. {IEEE}
  Glob. Commun. ({GLOBECOM})}, Singapore, Singapore, Dec. 2017, pp. 1--7.

\bibitem{URLLC-Pilot}
H.~{Ren}, C.~{Pan}, Y.~{Deng}, M.~{Elkashlan}, and A.~{Nallanathan}, ``Joint
  pilot and payload power allocation for massive-{MIMO}-enabled {URLLC} {IIoT}
  networks,'' \emph{{IEEE} J. Sel. Areas Commun.}, vol.~38, no.~5, pp.
  816--830, Mar. 2020.

\bibitem{URLLC-MIMO-TVT-2019}
J.~{Zeng}, T.~{Lv}, R.~P. {Liu}, X.~{Su}, N.~C. {Beaulieu}, and Y.~J. {Guo},
  ``Linear minimum error probability detection for massive {MU-MIMO} with
  imperfect {CSI} in {URLLC},'' \emph{{IEEE} Trans. Veh. Technol.}, vol.~68,
  no.~11, pp. 11\,384--11\,388, Nov. 2019.

\bibitem{URLLC-MIMO-IOT-2020}
J.~{Zeng}, T.~{Lv}, R.~P. {Liu}, X.~{Su}, Y.~J. {Guo}, and N.~C. {Beaulieu},
  ``Enabling ultrareliable and low-latency communications under shadow fading
  by massive {MU-MIMO},'' \emph{{IEEE} Internet Things J.}, vol.~7, no.~1, pp.
  234--246, Jan. 2020.

\bibitem{URLLC-MIMO-IOT-RA-2019}
J.~{Ding}, D.~{Qu}, H.~{Jiang}, and T.~{Jiang}, ``Success probability of
  grant-free random access with massive {MIMO},'' \emph{{IEEE} Internet Things
  J.}, vol.~6, no.~1, pp. 506--516, Feb. 2019.

\bibitem{URLLC-2018-CBA}
B.~Singh, O.~Tirkkonen, Z.~Li, and M.~A. Uusitalo, ``Contention-based access
  for ultra-reliable low latency uplink transmissions,'' \emph{{IEEE} Wireless
  Commun. Lett.}, vol.~7, no.~2, pp. 182--185, Apr. 2018.

\bibitem{URLLC-MIMO-RE-2019}
R.~{Kotaba}, C.~N. {Manchón}, and P.~{Popovski}, ``Enhancing performance of
  uplink {URLLC} systems via shared diversity transmissions and multiple
  antenna processing,'' in \emph{2019 53rd Asilomar Conference on Signals,
  Systems, and Computers}, Pacific Grove, CA, USA, Nov. 2019, pp. 1409--1415.

\bibitem{Linlin-urllc-CL}
L.~Zhao, X.~Chi, L.~Qian, and W.~Chen, ``Analysis on latency-bounded
  reliability for adaptive grant-free access with multipackets reception
  ({MPR}) in {URLLCs},'' \emph{{IEEE} Commun. Lett.}, vol.~23, no.~5, pp.
  892--895, May 2019.

\bibitem{EE-WCNC-2018}
R.~Abreu, T.~Jacobsen, G.~Berardinelli, K.~Pedersen, I.~Kovacs, and
  P.~Mogensen, ``Power control optimization for uplink grant-free {URLLC},'' in
  \emph{Proc. {IEEE} Wireless Commun. Netw. Conf. ({WCNC})}, Barcelona, Spain,
  Apr. 2018, pp. 1--6.

\bibitem{Linlin-martingales-URLLC}
L.~Zhao, X.~Chi, and Y.~Zhu, ``Martingales-based energy-efficient {D-ALOHA}
  algorithms for {MTC} networks with delay-insensitive/{URLLC} terminals
  co-existence,'' \emph{{IEEE} Internet Things J.}, vol.~5, no.~2, pp.
  1285--1298, Apr. 2018.

\bibitem{URLLC-EE-ICCC2019}
X.~{Tang}, Z.~{Zhang}, L.~{Wu}, and J.~{Dang}, ``Energy efficient fuzzy time
  slot scheduling scheme for massive connections,'' in \emph{Proc. {IEEE/CIC}
  Int. Conf. Commun. in China ({ICCC})}, Changchun, China, Aug. 2019, pp.
  117--122.

\bibitem{URLLC-EE-DU}
C.~{Sun}, C.~{She}, C.~{Yang}, T.~Q.~S. {Quek}, Y.~{Li}, and B.~{Vucetic},
  ``Optimizing resource allocation in the short blocklength regime for
  ultra-reliable and low-latency communications,'' \emph{{IEEE} Trans. Wireless
  Commun.}, vol.~18, no.~1, pp. 402--415, Nov. 2019.

\bibitem{URLLC-2017-ICC}
G.~Pocovi, B.~Soret, K.~I. Pedersen, and P.~Mogensen, ``{MAC} layer
  enhancements for ultra-reliable low-latency communications in cellular
  networks,'' in \emph{Proc. {IEEE} Int. Conf. Commun. ({ICC})}, Paris, France,
  May 2017, pp. 1005--1010.

\bibitem{TI_PROCEEDING_2019}
E.~{Steinbach}, M.~{Strese}, M.~{Eid}, X.~{Liu}, A.~{Bhardwaj}, Q.~{Liu},
  M.~{Al-Ja’afreh}, T.~{Mahmoodi}, R.~{Hassen}, A.~{El Saddik}, and
  O.~{Holland}, ``Haptic codecs for the tactile internet,'' \emph{Proc.
  {IEEE}}, vol. 107, no.~2, pp. 447--470, Feb. 2019.

\bibitem{open-loop-power-control}
Y.~{Liu}, Y.~{Deng}, M.~{Elkashlan}, A.~{Nallanathan}, and G.~K.
  {Karagiannidis}, ``Analyzing grant-free access for {URLLC} service,''
  \emph{{IEEE} J. Sel. Areas Commun.}, vol.~39, no.~3, pp. 741 -- 755, Mar.
  2021.

\bibitem{ZF-channel-error}
C.~{Wang}, E.~K.~S. {Au}, R.~D. {Murch}, W.~H. {Mow}, R.~S. {Cheng}, and
  V.~{Lau}, ``On the performance of the {MIMO} zero-forcing receiver in the
  presence of channel estimation error,'' \emph{{IEEE} Trans. Wireless
  Commun.}, vol.~6, no.~3, pp. 805--810, Mar. 2007.

\bibitem{URLLC-proceeding}
G.~Durisi, T.~Koch, and P.~Popovski, ``Toward massive, ultrareliable, and
  low-latency wireless communication with short packets,'' \emph{Proc. {IEEE}},
  vol. 104, no.~9, pp. 1711--1726, Sep. 2016.

\bibitem{channel-coding}
Y.~Polyanskiy, H.~V. Poor, and S.~Verdu, ``Channel coding rate in the finite
  blocklength regime,'' \emph{{IEEE} Trans. Inf. Theory}, vol.~56, no.~5, pp.
  2307--2359, May 2010.

\bibitem{URLLC-magzine}
C.~She, C.~Yang, and T.~Q.~S. Quek, ``Radio resource management for
  ultra-reliable and low-latency communications,'' \emph{{IEEE} Commun. Mag.},
  vol.~55, no.~6, pp. 72--78, Jun. 2017.

\bibitem{URLLC-Secure}
H.~{Ren}, C.~{Pan}, Y.~{Deng}, M.~{Elkashlan}, and A.~{Nallanathan}, ``Resource
  allocation for secure {URLLC} in mission-critical {IoT} scenarios,''
  \emph{{IEEE} Trans. Commun.}, vol.~68, no.~9, pp. 5793--5807, Jun. 2020.

\bibitem{MPR-Access-YunH2014}
Y.~Bae, B.~Choi, and A.~Alfa, ``Achieving maximum throughput in random access
  protocols with multipacket reception,'' \emph{{IEEE} Trans. Mobile Comput.},
  vol.~13, no.~3, pp. 497--511, Mar. 2014.

\bibitem{Markov}
J.~G. Kemény and J.~L. Snell, \emph{Finite Markov Chains (Second ed.)}.\hskip
  1em plus 0.5em minus 0.4em\relax New York Berlin Heidelberg Tokyo:
  Springer-Verlag, 1976.

\bibitem{MRC-SINR}
G.~{Zhu}, C.~{Zhong}, H.~A. {Suraweera}, Z.~{Zhang}, and C.~{Yuen}, ``Outage
  probability of dual-hop multiple antenna {AF} systems with linear processing
  in the presence of co-channel interference,'' \emph{{IEEE} Trans. Wireless
  Commun.}, vol.~13, no.~4, pp. 2308--2321, Mar. 2014.

\bibitem{URLLC-Yang-crosslayer}
C.~She, C.~Yang, and T.~Q.~S. Quek, ``Cross-layer optimization for
  ultra-reliable and low-latency radio access networks,'' \emph{{IEEE} Trans.
  Wireless Commun.}, vol.~17, no.~1, pp. 127--141, Jan. 2018.

\bibitem{6G-learning}
C.~She, C.~Sun, Z.~Gu, Y.~Li, C.~Yang, H.~V. Poor, and B.~Vucetic, ``A tutorial
  on ultra-reliable and low-latency communications in {6G}: Integrating domain
  knowledge into deep learning,'' \emph{Proc. {IEEE}}, vol. 109, no.~3, pp.
  204--246, Mar. 2021.

\bibitem{book-MST}
R.~J. Muirhead, \emph{Aspects of Multivariate Statistical Theory}.\hskip 1em
  plus 0.5em minus 0.4em\relax New York, NY: USA: Wiley, 1982.

\end{thebibliography}

%

\begin{IEEEbiography}[{\includegraphics[width=1in,height=1.25in,clip,keepaspectratio]{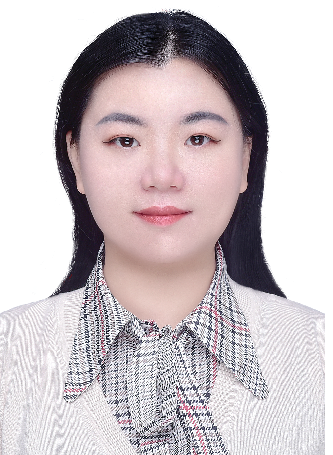}}]{Linlin Zhao}
 received the B.Eng., M.S., and Ph.D. degrees from the Department of Communications Engineering, Jilin University, Changchun, China, in 2009, 2012, and 2017, respectively. From 2017 to 2019, she held a postdoctoral position at the Department of Communications Engineering, Jilin University. Since September 2019, she has been with Jilin University, where she is currently an Associate Professor. Currently she is also a Postdoctoral Research Fellow with the State Key Laboratory of Internet of Things for Smart City, University of Macau, Macau, China. Her current research interests include throughput optimal random access algorithms, resource allocation schemes, as well as delay/reliability analysis and optimization, especially for reliability analysis of ultra-reliable low-latency communications. She was a recipient of the Best Ph.D. Thesis Award of Jilin University in 2017, and an awardee of the Macau Young Scholars Program in 2019.
\end{IEEEbiography}

\begin{IEEEbiography}[{\includegraphics[width=1in,height=1.25in,clip,keepaspectratio]{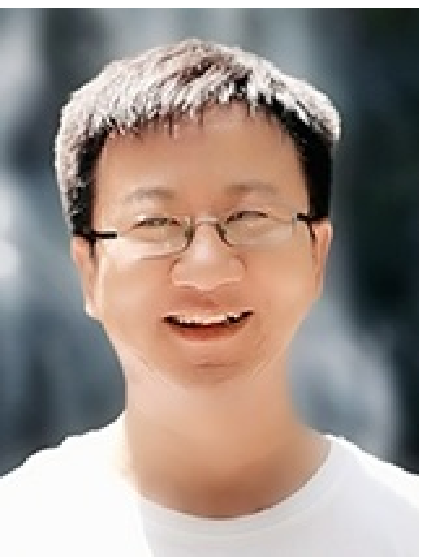}}]{Shaoshi Yang}
	(S’09–M’13–SM’19) received the B.Eng. degree in information engineering from the Beijing University of Posts and Telecommunications (BUPT), China, in 2006, and the Ph.D. degree in electronics and electrical engineering from the University of Southampton, U.K., in 2013. From 2008 to 2009, he was a researcher of WiMAX standardization with Intel Labs China. From 2013 to 2016, he was a Research Fellow with the School of Electronics and Computer Science, University of Southampton. From 2016 to 2018, he was a Principal Engineer with Huawei Technologies Co. Ltd., where he made substantial contributions to the company’s products and solutions on 5G base stations, wideband IoT, and cloud gaming/VR. He is currently a Full Professor with BUPT. His research expertise includes 5G wireless networks, massive MIMO, iterative detection and decoding, mobile ad hoc networks, distributed artificial intelligence, and cloud gaming/VR. He is a member of the Isaac Newton Institute for Mathematical Sciences, Cambridge University, and a Senior Member of IEEE. He received the Dean’s Award for Early Career Research Excellence from the University of Southampton in 2015, the Huawei President Award of Wireless Innovations in 2018, the IEEE Technical Committee on Green Communications \& Computing (TCGCC) Best Journal Paper Award in 2019, and the IEEE Communications Society Best Survey Paper Award in 2020. He is an Editor for IEEE Systems Journal, IEEE Wireless Communications Letters, and Signal Processing (Elsevier). He was also an invited international reviewer of the Austrian Science Fund (FWF).
\end{IEEEbiography}

\begin{IEEEbiography}[{\includegraphics[width=1in,height=1.25in,clip,keepaspectratio]{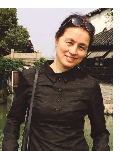}}]{Xuefen Chi} 
	received the B.Eng. degree in applied physics from the Beijing University of Posts and Telecommunications, Beijing, China, in 1984, and the M.S. and Ph.D. degrees from the Changchun Institute of Optics, Fine Mechanics and Physics, Chinese Academy of Sciences, Changchun, China, in 1990 and 2003, respectively. She was a Visiting Scholar with the Department of Computer Science, Loughborough University, U.K., in 2007, and the School of Electronics and Computer Science, University of Southampton, Southampton, U.K., in 2015. She is currently a Professor with the Department of Communications Engineering, Jilin University, China. Her research interests include machine-type communications, indoor visible light communications, random access algorithms, delay-QoS guarantees, and network modelling theory and its applications.
\end{IEEEbiography}

\begin{IEEEbiography}[{\includegraphics[width=1in,height=1.25in,clip,keepaspectratio]{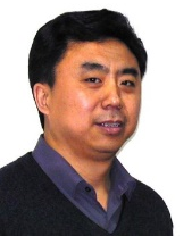}}]{Wanzhong Chen}
	received his Ph.D. degree in power machinery and engineering from Jilin University, Jilin, China, in 2001. Since 2006, he has been a Professor with Jilin University. His research interests include intelligent signal processing, brain–computer interface and pattern recognition.
\end{IEEEbiography}

\begin{IEEEbiography}[{\includegraphics[width=1in,height=1.25in,clip,keepaspectratio]{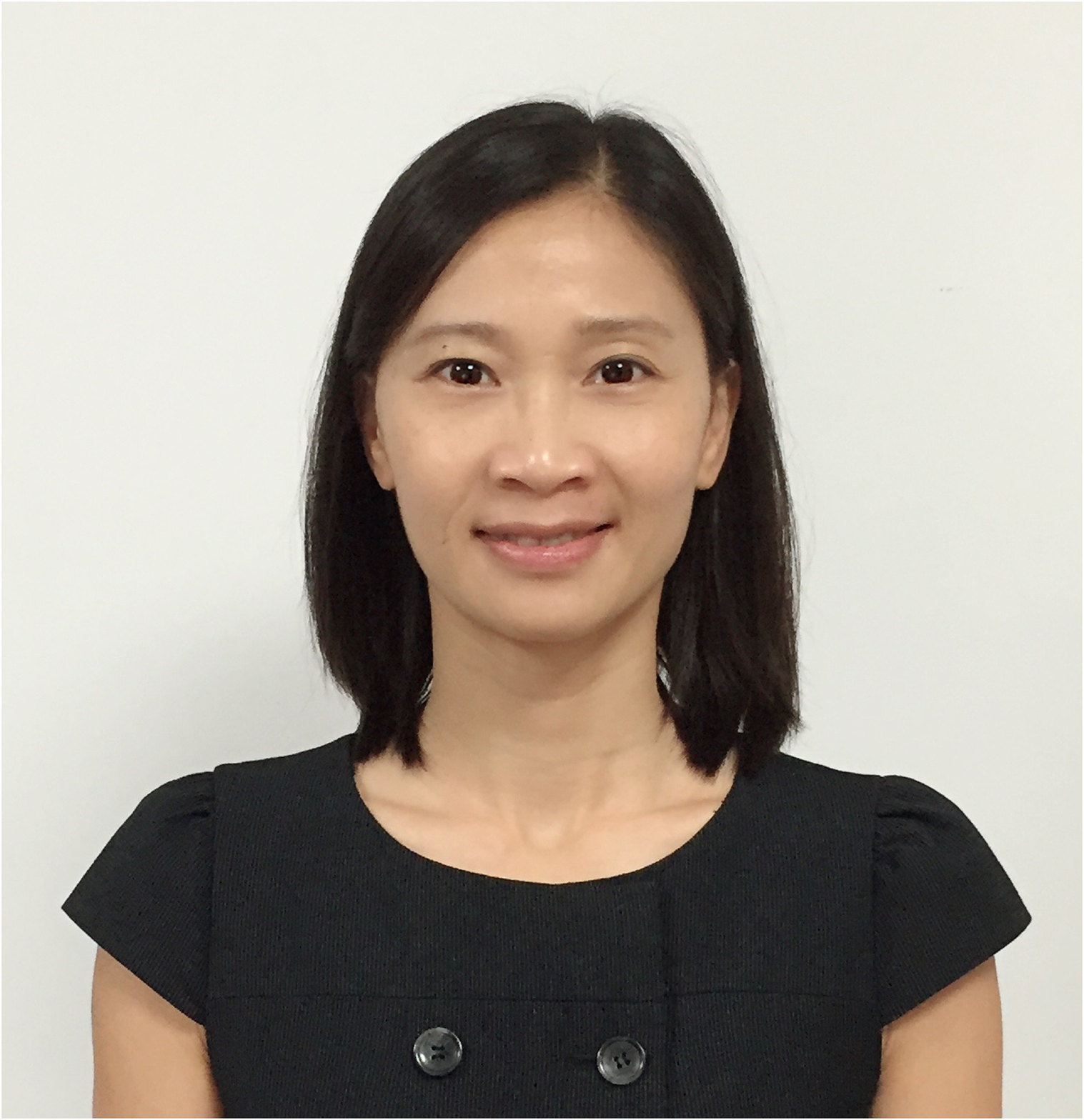}}]{Shaodan Ma}
	received the double bachelor’s degrees in science and economics and the master’s degree in engineering from Nankai	University, Tianjin, China, in 1999 and 2002, respectively, and the Ph.D. degree in electrical and electronic engineering from the University of Hong	Kong, Hong Kong, in 2006. From 2006 to 2011, she was a Postdoctoral Fellow with the University of Hong Kong. Since August 2011, she has been with the University of Macau,	Macau, China, where she is currently an Associate Professor. She was a Visiting Scholar with Princeton University, Princeton, NJ, USA, in 2010. Her research interests are in the general areas of signal processing and communications, particularly, transceiver design, resource allocation, and performance analysis.
\end{IEEEbiography}

%
%
%




\end{document}